%% file: main.tex
\RequirePackage{fix-cm}
\documentclass[smallextended]{svjour3}       % onecolumn (second format)
\smartqed  % flush right qed marks, e.g. at end of proof
\usepackage{graphicx}
\usepackage{cite}
\usepackage{amsmath,amssymb,amsfonts}
\usepackage{algorithmic}
\usepackage[ruled,vlined]{algorithm2e}
\usepackage{graphicx}
\usepackage{textcomp}
\usepackage{xcolor}
\usepackage{url}
\usepackage{footnote}
\usepackage[perpage]{footmisc}
\usepackage{balance}
\usepackage{multirow}
\usepackage{listings}

\usepackage{tcolorbox}

\usepackage[english=american]{csquotes}
\usepackage[authoryear,round]{natbib}

\usepackage{caption}
\usepackage{subcaption}
\graphicspath{ {./} }
\usepackage{array}
\newcommand{\al}{\textit{et al. }}

\newcommand{\practitioner}[1]{{\it[#1]}}
\newcommand{\issue}[1]{{\it[#1]}}
\newcommand{\citepr}[1]{\textit{``#1"}}

\journalname{EMSE 2023}
\begin{document}
\input{title_page}

\section{Introduction}
Nowadays, Machine Learning Software Systems (MLSSs) have become a part of our daily life (e.g. recommendation systems, speech recognition, face detection). Increasing demand is observed in various companies to employ Machine Learning (ML) for solving problems in their business. Typically, an MLSS receives data as input and employs ML models to make intelligent decisions automatically based on learned patterns, associations, and knowledge from data \citep{marijan2019challenges}. Therefore, ML models are implemented as software components integrated into other subsystems in MLSSs, and like other software systems, quality assurance is necessary. According to the growing importance of MLSSs in today’s world, there is a strong need for ensuring their quality. False or poor decisions of such systems can lead to the malfunction of other systems, significant financial losses, or even threats to human life \citep{foidl2019risk}. Recently, an automated vehicle caused a crash of 8 vehicles, leading to 2 juveniles being transported to a hospital \citep{Guardian_2022}. As reported by United States safety regulators, this is not a punctual event, since more than 400 car-crash with automated vehicles have been reported between July 2021 and May 2022 \citep{Krisher_2022}. Unqualified MLSSs can be harmful to society in less obvious ways. For example, an experimental job recruitment tool developed by Amazon was shown to reject women's job applications more often than men's \citep{Dastin_2018}. ML models acted unfairly in many other critical situations \citep{Martinez_Kirchner_2021, compass}. In addition to the dire consequences caused by unreliable or unfair models, large ML models often have a significant ecological footprint. For example, the energy needed to train the MegatronLM model \citep{shoeybi2019megatron} developed by Nvidia was almost equal to the annual consumption of three households for a year \citep{Labbe_2021}. Considering the growing popularity of ML, addressing the quality issues in MLSSs becomes of increasing importance. However, to solve an issue, one must first understand it. While there are research works describing the challenges in MLOps \citep{paleyes2022challenges, shankar2022operationalizing, steidl2023pipeline, 10.1145/3035918.3054782, polyzotis2018data, schelter2015challenges} and (anti-) patterns in MLSSs \citep{van2021prevalence, dilhara2021understanding, alahdab2019empirical, mailachsocio, bogner2021characterizing, washizaki2019studying, sambasivan2021everyone}, we have not observed a large literature focused on discovering and understanding the quality issues themselves.

In this paper, we investigate the characteristics of real-world quality issues in MLSSs from the viewpoint of practitioners. This is a requirement of comprehensive quality assessment of MLSSs, as Zhang \al already acknowledged the lack of such empirical study and asserted that conducting empirical studies on the prevalence of poor models among deployed ML models should be interesting \citep{zhang2020machine}. This study attempts to cover all relevant quality factors like performance (accuracy), robustness, explainability, efficiency, maintainability, and reliability. We conduct a set of interviews with practitioners/experts, to gather insights about their experience and practices when dealing with ML quality issues.

In total, we interviewed 42 practitioners; 14 of them are employees of MoovAI. The others were either found over GitHub using its API \footnote{\url{https://docs.github.com/en/rest?apiVersion=2022-11-28}} (26 practitioners) or were personal contacts (1 practitioner). To analyze the results, we performed a round of open coding, followed by a round of axial coding \citep{strauss1994grounded}. As a result, we extracted a list of 18 quality issues with MLSSs, for which we explain the causes and consequences. We report the current mitigation techniques used by the practitioners to address the identified quality issues. In addition to the quality issues with MLSSs, we share the data quality issues experienced by practitioners. To better understand them, we describe the data quality aspects that are the most problematic in practice according to the practitioners' experience. We also explain the challenges that occur with different data collection processes that lead to data quality issues. To validate the identified quality issues with MLSSs, we conducted a survey with practitioners found on social networks or from our personal contacts. In the survey questionnaire, the practitioners were asked to rate on a five-level Likert scale how often they are faced with the quality issues we found. In total, we obtained 21 answers. Our findings show that (1) using ML in software systems brings a unique set of challenges, and (2) practitioners struggle to build qualified MLSSs. Most of the issues stem from (1) a lack of appropriate tooling, (2) organizational challenges, or (3) limited knowledge of a topic among developers/practitioners. Through this study, we aim to contribute knowledge that will help develop efficient quality assurance tools for ML models and MLSSs. A replication package of our study is available on our public GitHub repository\footnote{\url{https://github.com/poclecoqq/quality_issues_in_MLSSs}}.

\vspace{1em}
\noindent \textbf{Contributions}. In summary, we make the following contributions. 

\begin{itemize}
    \item We describe 18 quality issues with MLSSs identified during our interviews. For each quality issue, we explain its causes and its consequences from the point of view of the practitioners. We also inform which ones are the most prevalent in practice.
    \item We share 21 strategies used by the practitioners we interviewed to mitigate the aforementioned quality issues. We also propose 12 other strategies to address the issues based on the literature and the root causes of the issues.
    \item In addition to the quality issues in MLSSs, we describe the data quality issues encountered by our practitioners. We show how frequent they are according to the practitioners' experience. 
    \item We describe the challenges encountered with different data collection processes that lead to data quality issues. 
    \item  We characterize the challenges of data quality assurance during model evolution. 
    \item Based on the quality issues (with MLSSs and with data) we found, we provide 9 future work recommendations.
\end{itemize}

\vspace{1em}
\noindent \textbf{The rest of this paper is organized as follows}. Section \ref{relatedwork} reviews relevant related work to our study. In Section \ref{metho}, we define our research questions and our methodology. Section \ref{results} presents the results of our survey. Finally, in Section \ref{recommendations}, we provide recommendations to practitioners (regarding handling quality issues with data and MLSSs) and to researchers (for future works).

\section{Related Work}\label{relatedwork}
We organize the related works of our study into four groups. The first step toward building a qualified MLSS is defining necessary requirements. Hence, we start by presenting related works in that area. The following two parts are dedicated to topics tightly related to quality issues: technical debt and anti-patterns, and general challenges in MLOps. The final group discusses studies that propose potential solutions to quality issues.

\subsection{Requirement Engineering}
One of the first steps when building a software system is to define good requirements. These requirements specify functional and non-functional objectives that must be obtained for the MLSS to be considered a success. A system that does not achieve these goals will most likely have quality issues. \citet{vogelsang2019requirements} describes some of the additional challenges MLSSs have compared to traditional software systems during requirement engineering. Notably, new quality aspects must be considered when defining requirements for MLSSs, such as explainability, fairness, and legal aspects of data. Similar work is done by \citet{10098194, horkoff2019non}. \citet{horkoff2019non} focused on non-functional requirements and explained the difficulty of defining measurable success criteria for quality attributes. For example, how should maintainability be measured for an ML system? In reaction to the aforementioned issue, \citet{siebert2022construction} proposes an approach to objectively specify and assess the quality attributes of MLSSs.

\subsection{Technical Debt and Anti-Patterns in MLSSs}
\citet{avgeriou2016managing} defines Technical Debt (TD) as `` [any] design or implementation constructs that are expedient in the short term, but set up a technical context that can make a future change more costly or impossible''. Similarly, anti-patterns refer to recurring solutions to a problem that leads to sub-optimal results \citep{bogner2021characterizing}. While both were initially conceptualized for systems without ML components, recent studies transpose these concepts to MLSSs. For example, the seminal work of  \citet{NIPS2015_86df7dcf} shared the challenges and types of TD faced by Google while building, deploying, and maintaining ML models. If left unattended, TD and anti-patterns in an MLSS may result in quality issues. For example, a model with undeclared consumers (the anti-pattern) may hinder the reliability of an MLSS (the quality issue) \citep{bogner2021characterizing}. In the following, we present other studies that try to uncover the types of technical debt and anti-patterns commonly found in MLSSs. We also present studies on code smells, since code smells are tightly connected to anti-patterns. Code smells are potential indicators of bad practices, while anti-patterns are a bad practice \citep{bogner2021characterizing}.

\citet{van2021prevalence} mined open-source GitHub repositories to aggregate a list of the most common maintenance-related modifications in Deep Learning (DL) projects. From this list, they extracted 5 code smells in DL systems. Furthermore, they measured how prevalent and problematic the code smells are from the point of view of practitioners using a survey. Similarly, \citet{dilhara2021understanding} mined 26 open-source MLSSs and identified 14 refactorings and 7 new TD categories specific to ML. Instead of finding new types of TD, \citet{alahdab2019empirical} shared with the scientific community how it appeared in the early phases of an industrial DL project. \cite{nikanjam2021design} identified and cataloged 8 design smells that frequently occur in DL systems by reviewing the existing literature on DL systems' design and manually inspecting 659 DL programs with performance issues and design inefficiencies. On the same line of work, \citet{mailachsocio} analyzed 66 hours of talks recorded by an MLOps community to extract a list of 17 socio-technical anti-patterns. Finally, \cite{Jebnoun22} examined the frequency, distribution, and impacts of code clones in DL systems, as well as developers' code cloning practices. Facing the increasing number of papers regarding technical debt and code smells in MLSSs, \citet{bogner2021characterizing} attempted to aggregate the knowledge on these topics by performing a systematic mapping study that presented 4 new types of technical debt, 72 anti-patterns along with 46 solutions. Similar work has been done by \citet{washizaki2019studying} but by consulting gray literature as well. \citet{sambasivan2021everyone} describes different challenges with data causing negative downstream effects on MLSSs such as technical debt. The main difference between these works and ours is that they are focused on the causes of issues (anti-patterns, technical debts, code smells), while we are focusing on the issues themselves. Furthermore, we are only interested in identifying ML quality issues and potential strategies to address them.

\subsection{Challenges in MLOps}
In the following, we present papers discussing the general topics of challenges in developing and deploying MLSSs. These challenges are often tightly coupled with an underlying quality issue. For example, \citet{paleyes2022challenges} mentions that the hyper-parameter selection is resource-intensive (challenge), which implies an issue with efficiency. Similar to our project, \citet{shankar2022operationalizing} conducted an interview study with practitioners to discover their pain points along with common anti-patterns. They also provide a set of best practices for successful MLSS development and maintenance. \citet{paleyes2022challenges} performed a survey on the challenges when building and deploying MLSSs. These challenges sometimes affect quality aspects, such as maintainability, explainability, and fairness. Similar to the works above, \citet{steidl2023pipeline} listed the challenges experienced by practitioners at different stages of the development of an MLSS. To do so, they performed a Multivocal Literature Review and verified their findings through interviews with practitioners. Other works focused on the challenges of specific aspects of an MLSS. \citet{10.1145/3035918.3054782, polyzotis2018data} discuss data management challenges and \citet{schelter2015challenges} presents model management challenges. The main difference between these works and ours is the scope. We are only interested in quality issues, while the aforementioned studies are interested in any recurring challenge in MLOps.

\subsection{Tools and Techniques to Address Quality Issues}
In reaction to the growing concern about quality issues in MLSSs, researchers have proposed techniques and tools to help practitioners develop qualified systems. One must define quality beforehand to detect quality issues. Therefore, for data, \citet{cappi2021dataset} describes different aspects that must be considered when evaluating the quality of a dataset. Some examples of quality aspects are data representativeness or data accuracy and precision. Other tools such as the DataLinter \citep{hynes2017data} or the data validation component of TFX \citep{breck2019data} have been created to automate data quality assurance. Another group of researchers adopted a different approach and instead shared a checklist of tests to assess the production readiness of MLSSs \citep{breck2017ml}. A similar work is done by \citet{fujii2020guidelines}. \citet{amershi2019software} presents a list of best practices when developing MLSSs. In a similar fashion, \citet{studer2021towards, lavin2022technology} and TDSP\footnote{\url{https://learn.microsoft.com/en-us/azure/architecture/data-science-process/overview}} propose a process for the development of ML applications where quality assurance plays a central role. Finally, \citet{berglundtest} identifies factors that make test maintenance difficult in MLSSs and proposes ten recommendations to make it easier.

\section{Study Design} \label{metho}
\begin{figure}[t]
 \centering
 \includegraphics[width=1\textwidth]{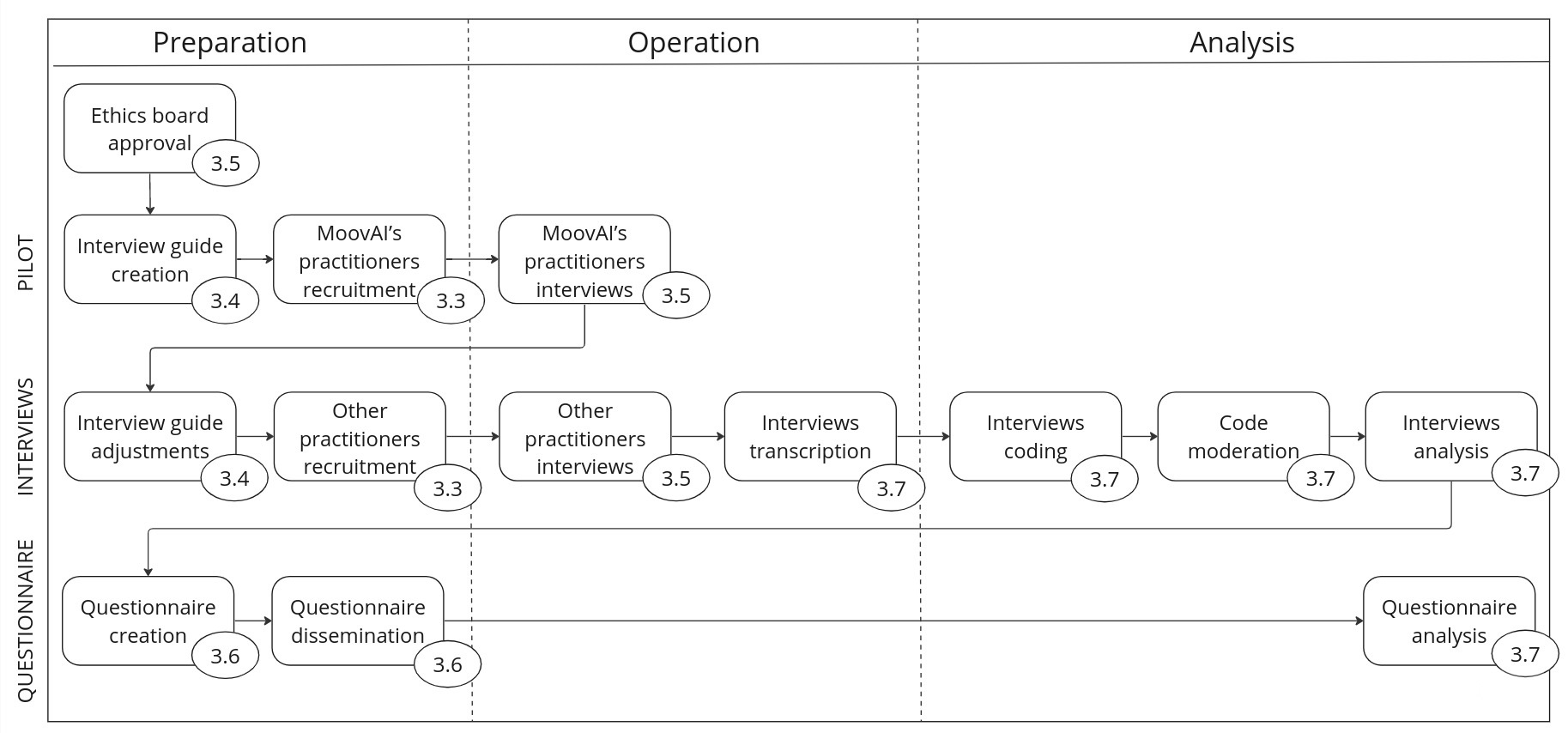}
 \caption{Overview of the methodology of our study.}
 \label{fig:metho_overview}
\end{figure}

In this section, we describe how we conducted this study. An overview of our methodology is given in Figure \ref{fig:metho_overview}. We indicate the section in the paper (e.g. 3.3) where more information can be obtained for a particular activity in the figure. In the following, we start by presenting the research questions that guided our study. Then, we introduce our industrial partners who helped us conduct this study. Finally, we describe every step we followed to complete this study.

\subsection{Objectives and Research Questions}
The goal of this study is to provide a detailed analysis of quality issues (including data and model) in MLSSs. We believe that interviewing people who really experienced these issues is an effective way to gain that knowledge. Thus, we define the following Research Questions (RQ):
\begin{description}
    \item[RQ1:] \textbf{What are the quality issues encountered by practitioners when building MLSSs?} For future works to solve quality issues, they must know the issues that exist. While there is some literature covering some quality issues \citep{NIPS2015_86df7dcf, nahar2022collaboration,studer2021towards}, we believe that a lot is still unknown. In this study, we aim to share with the research community the quality issues encountered by practitioners when building ML systems. This includes issues related to data, models, and other components in MLSSs. We describe the \textbf{cause} and \textbf{consequences} of the discovered quality issues according to practitioners' experience.
    
    \item[RQ2:] \textbf{Which quality issues are the most prevalent?} To help practitioners focus on the quality issues that are experienced more frequently, we also measure how frequent the discovered issues are in practitioners' experience.
    
    \item[RQ3:] \textbf{How are the quality issues currently handled by the practitioners?} Once quality issues have been detected (e.g., in data or model), we expect practitioners to have put in place mechanisms to mitigate or at least attenuate their consequences. We are interested in understanding the current mitigation approaches implemented in the industry. 

    \item[RQ4:] \textbf{Which data quality aspect is the most problematic in practice?} As it is widely known in the ML community, a large part of the practitioners' time and effort when building ML solutions is spent on data \citep{whang2021data}. Recent works \citep{cappi2021dataset, dama} have tried to build lists of desirable properties of data for ML (e.g. correct labels), to help create a common vocabulary to describe data quality. In the following, we will refer to these desirable data properties as data quality aspects. Hence, using \citet{cappi2021dataset}'s taxonomy, we aim to identify the data quality aspects that are the most frequently impacted by the data quality issues experienced by practitioners.   
    
    \item[RQ5:] \textbf{In the case of data, which data types and collection processes are the most challenging in terms of data quality?} The training data ingested by ML models come in many forms. Face recognition systems use images, while stock prediction applications can use numbers/amounts (like time-series data) or even text. Each data type comes with its own data quality challenges when training ML models. By answering this RQ, future research will be guided toward the most pressing data quality challenges faced by practitioners. Authors in \citep{whang2021data} mention that data collection processes, the process of gathering data, may affect the quality of data. For example, one could choose to train one's model on public datasets, or manually acquire more data with data collectors. Each of these processes has different challenges which may lead to different data quality issues. In this study, we want to identify the data collection processes that are most prone to data quality issues.
    
    \item[RQ6:] \textbf{What are the challenges of data quality assurance during model evolution?}    
    Many MLSSs encounter rapidly changing/non-stationary data, adversarial input, or differences in data distribution (concept drift, e.g., content recommendation systems and financial ML applications). Hence, the quality of the deployed model may decrease over time and consequently affects the performance of the whole system. Therefore, the robustness and the accuracy of the model's predictions must be assessed frequently in production. Actively monitoring the quality of the deployed model in production is crucial to detect performance degradation and model staleness.
        
\end{description}

\subsection{Industrial Partner}
MoovAI\footnote{\url{https://moov.ai/en/}} is a Montreal-based company in Canada, which is active in developing AI/ML quality assurance solutions to address practical needs in various businesses. MoovAI’s experts guide their customers (i.e., companies) to take advantage of these cutting-edge technologies, regardless of their level of maturity in data science. Nowadays, various sectors in industry are and will be developing ML models in their systems, e.g., the energy sector, the financial sector, supply chain recommendations, medical diagnosis, and treatment. As ML-based technologies become more widespread, the quality demands of ML become more important. Poor quality models are a potential barrier to exploiting ML in real-world applications and currently more companies request qualified MLSSs \citep{CD4MLSato}. A study showed that 87\% of ML proof of concepts have never come to production \citep{azimi2020root}, usually due to lack of enough quality.

Currently, MoovAI is using a model validation tool to validate ML models prior to deployment in the real world \citep{Blais2020}. This tool includes preliminary validation methods to assess the overall quality of an ML model. The tool evaluates the accuracy, stability, biases, and sensitivity of ML models. Now, MoovAI aims to push its tool forward to develop advanced methods for ensuring the quality of not only the ML models but also ML-based systems (i.e., software systems containing ML components) and to develop a stand-alone model validation platform. Experts at MoovAI now are looking for a comprehensive quality evaluation tool to assess and monitor the quality of ML models during their whole life cycle; from data collection to development, deployment, and maintenance. 

\subsection{Participants}  \label{section:participants}
In this section, we describe how we selected candidates for interviews and how we recruited them. In total, \textbf{we interviewed 42 participants}, which is more than previous similar studies \citep{humbatova2020taxonomy, serban2021empirical}.

\subsubsection{Inclusion and exclusion criteria} \label{participants:inclusion-exclusion}
In the following, we list the inclusion and exclusion criteria used to select practitioners to interview. In most cases, we could apply the exclusion criteria only after the interview had been conducted and we knew the practitioner’s experience Hence, after conducting the interviews, we excluded 6 of them, \textbf{leaving 36 interviews for further analysis}. We excluded 3 participants from our study because they did not have any experience working on MLSSs, 2 others because we could not understand them because of a language barrier, and 1 because the interviewee had less than a year of experience with ML.

\smallskip
\underline{Inclusion criteria}:
\begin{itemize}
    \item The interviewee is a data scientist or a machine learning engineer.
    \item The interviewee is a software engineer, a project manager, or a data engineer and has worked on MLSSs.
\end{itemize}

\underline{Exclusion criteria}:
\begin{itemize}
    \item The interviewee has not worked on MLSSs.
    \item The interviewee has less than a year of experience in ML.
    \item The interviewee can not be understood by the interviewers because of a language barrier and the tool used to transcribe the interview is not able to provide an intelligible transcript.
\end{itemize}

\subsubsection{Participant recruitment} \label{particpants:recruitment}
We conducted two rounds of interviews. In the first round of the study, we interviewed the employees of our industrial partner (i.e., MoovAI). They are Data Scientists, Machine Learning Engineers, Data Engineers, and Project Managers (of ML projects). They have worked on many different ML projects for different clients. Thus, we think that they are able to provide a global view of quality issues encountered in the industry. We interviewed them between July 2022 and September 2022. In total, we interviewed 14 employees from MoovAI. In the second phase of the study, we interviewed practitioners from other companies. We interviewed them between November 2022 and February 2023. We adopted 4 strategies to recruit participants for the second round of interviews.
\begin{itemize}
    \item \textit{Q\&A websites}: Questions and Answers websites are platforms on which a lot of knowledge is shared. Following similar previous work \citep{humbatova2020taxonomy}, we searched for practitioners with meaningful experience in ML willing to be interviewed on quality issues in ML systems. We chose to search on Stack Overflow\footnote{\url{https://stackoverflow.com/}} and Data Science Stack Exchange\footnote{\url{https://datascience.stackexchange.com/}} because both these websites are significantly used by the ML community for question answering. We reached out to the top askers and answerers\footnote{The words \textit{askers} and \textit{answerers} are part of the terminology used by the Q\&A websites to describe people asking and answering questions} since they have shown significant involvement in the ML community and most likely have expertise. Since Stack Overflow is also used by the Software Engineering community to ask questions on topics unrelated to ML, we only picked the top askers and answerers on topics related to quality issues in ML systems. Similar to \citet{humbatova2020taxonomy}, we selected the topics \textit{data-cleaning, dataset, machine-learning} and \textit{artificial-intelligence}. For each topic, the Q\&A websites provide the top askers and answerers of all time and the top askers and answerers of the last 30 days. We pick the top 10 practitioners from both lists. Hence, for each topic (Data Science Stack Exchange + 4 topics), we pick the top 10 askers and answerers from the \textit{`All Time'} and \textit{`Last 30 Days'} lists (10 users * 2 lists * 2 (askers and answerers) * 5 topics = 200 user profiles). 
    We manually searched for the users' email addresses using the information given in the user profile. In total, we obtained 44 email addresses, but \textbf{no practitioner} answered our invitation. We conducted that search in September 2022.

    \item \textit{GitHub}: GitHub is a hosting service for software development and version control using Git\footnote{\url{https://github.com/}}. Its users have a public user profile they can customize. We observed that some users chose to indicate in their bio their job role. Hence, we leveraged that observation and searched users who have the keywords ``Machine Learning Engineer" or ``Data Scientist" in their bio. We used a script available in our replication package to automatically search the users and fetch their email addresses using the GitHub API\footnote{\url{https://docs.github.com/en/rest?apiVersion=2022-11-28}}. In total, we managed to interview \textbf{27 practitioners} using this strategy.

    \item \textit{Social networks}: Social networks are platforms in which users may engage in conversations on a wide range of topics. Some of them host discussions on ML and Artificial Intelligence in general. We believe users with important ML expertise can be found on these websites. We post an invitation for interviews on the deep learning and ML communities of Reddit\footnote{\url{https://www.reddit.com/}} similar to our previous works \citep{nikanjam2021design}. The same invitation has been shared on two other MLOps communities on Slack\footnote{\url{https://mlops.community/}} and Discord\footnote{\url{https://mlops-discord.github.io/}}, similar to \citet{shankar2022operationalizing}. We posted our message in October 2022 and got \textbf{no answer}.
    
    \item \textit{Personal contacts}: We searched in our personal network for practitioners that have experience working with MLSSs. We managed to interview \textbf{1 practitioner}.
\end{itemize}

\subsection{Interview Structure} \label{metho:interview-design}
% Grounded theory
Since the topic of quality issues in MLSSs is not mature and we believe there is still room for research work, we follow the principles of a research procedure suited for exploratory work, Straussian Grounded Theory \citep{strauss1998basics}. It is a research method in which data collection and data analysis are executed in an iterative manner until a new theory emerges from the data. As opposed to many deductive approaches where a theory is first conceptualized and then tested through experiments, Grounded Theory goes the opposite way by inductively generating a theory from the data. Data collection stops when theoretical saturation is met: when the understanding of the subject is complete and new data does not invalidate the emerging theory. After 42 interviews, no new knowledge emerged from the interviews, so we stopped conducting interviews. Grounded Theory is often used when little is known about a phenomenon, because of its flexibility and its aptitude for the discovery of unknown concepts. Similar studies such as \citet{nahar2022collaboration} followed Grounded Theory. 

% Semi-structured interviews
Because we are doing exploratory work, the interviews have to be structured in a way that allows the interviewees to share knowledge we might not be aware of. For this reason, we conducted semi-structured interviews similar to \citet{humbatova2020taxonomy}. To help us cover every relevant topic, we devised an interview guide. It is composed mostly of open-ended questions so that the interviewee is able to direct us toward interesting information. It is the interviewer's responsibility to ask follow-up questions when the respondent touches upon a subject relevant to the study. To avoid the interviewer being overwhelmed with his tasks and having difficulty asking the right follow-up questions, each interview is conducted by two persons. This follows the recommendation of previous works which show that interviewees share more information when two interviewers are present rather than one \citep{1509301}. The role of the second interviewer is to help the primary interviewer to ask follow-up questions.

% Interview guide
To draft an initial set of questions for the interview guide, we considered the quality issues elicited by other studies \citep{NIPS2015_86df7dcf, serban2021empirical, bogner2021characterizing, washizaki2019studying, breck2017ml}. We also considered the data quality aspects defined by \citet{cappi2021dataset}. We completed the initial set of questions by drawing questions from a similar study \citep{humbatova2020taxonomy}. We assessed the quality of our interview guide by conducting a pilot of our study. We purposefully selected 5 participants with diverse experience (3 Data scientists, a Data engineer, and an AI project manager) and interviewed them using the interview guide. By doing so, we verified that our questions are unambiguous, precise, and able to answer our research questions effectively. The questions that did not satisfy the aforementioned criteria were improved. We also added questions we believed were missing in order to address our research questions.

\subsection{Interview Process} \label{metho:interview-process}
% Interviews
All the interviews were conducted in English except one that was conducted in French. After a round of introduction between the interviewee and the interviewer, we provide a brief overview of the project. Then we ask participants for some background information regarding their experience in ML. The interview officially starts with the following question: ``\textit{What are the main quality issues you have encountered with your data/model so far}"? By asking an open-ended question, we are allowing the interviewee to share experiences he is the most confident to talk about. Then we probe the interviewee’s experience in an attempt to discover quality issues. During the interview, we cover all the phases of ML workflows described by \citet{amershi2019software}, so that we can cover most situations where quality issues might occur. However, we do not cover the model requirement phase. For conciseness, we merged the data cleaning phase with the feature engineering one and the data labeling phase with the data collection one in our interview guide. We describe below the questions asked in every section of our interview.
\begin{itemize}
    \item \textit{Data collection}: We ask practitioners for their experience with data collection and the issues they have experienced. We also ask for the solutions they put in place to address the aforementioned issues.
    \item \textit{Data preparation}: Notably, we ask about pain points when preparing data for ML and for tools to automate the process. We consider any challenge related to the cleaning and transformation of the data. 
    \item \textit{Model evaluation}: We probe for issues the interviewee discovered when evaluating the quality of its model. For example, we ask the respondent if he was ever surprised by the poor performance of a model when it was evaluated by the users of an MLSS.
    \item \textit{Model deployment}: We gather general information about the process by which models are put into production. Then, we search for issues encountered at this step. For example, one question is: “\textit{Did you ever deploy a model that performed well locally but poorly once deployed?}”.
    \item \textit{Model maintenance}: We ask the interviewee how he ensures that the quality of its models remains the same after deployment. We specifically ask for past instances of model staleness and how it has been handled.
\end{itemize}
At the end of every section, we ask for any other issue that the interviewee may have at this step of the workflow in case we missed out on something. We conclude the interview with the open question: ``\textit{In your opinion, what is the most pressing quality issue researchers should try to solve?}". The answer to that question might provide interesting future work directions and, by asking the question, we follow Harvard's best practices for qualitative interviews\footnote{\url{https://sociology.fas.harvard.edu/files/sociology/files/interview_strategies.pdf}}. The interview guide is available in our replication package. 
 
A few days prior to the interview, we provide an overview of the interview with the participants so they can be familiar with the objectives of the study. Just before starting the interview, we ask the interviewees for permission to register and share the transcription. In order to follow ethical guidelines, we validated our research project with Research Ethics Committee at Polytechnique Montréal\footnote{\url{https://www.polymtl.ca/recherche/la-recherche-polytechnique/exigences-deontologiques/travaux-de-recherche-avec-des-etres-humains}} and got their approval. During interview coding, we clarify with the interviewee anything she mentioned that was ambiguous or could not be understood without further information.

\subsection{Questionnaire} \label{questionnaire}
After interviews have been conducted and analyzed, we expect to have a set of potential quality issues. In order to validate the quality of our findings and answer RQ2 (i.e., which quality issues are the most prevalent), we share a questionnaire with other practitioners. In this questionnaire, we present to the respondents our final list of quality issues and we ask them to rate on a scale from 1 to 5 how frequently they experience the described issue. Respondents have the choice of not answering a question if they think they do not have sufficient experience to answer the question. We also leave a free text field in the form where practitioners may write any comments. We collect demographic information at the beginning of the questionnaire. The questionnaire is built using Google Forms\footnote{\url{https://www.google.ca/forms/about/}}.

We used the same inclusion and exclusion criteria from Section \ref{participants:inclusion-exclusion} to select participants for our questionnaire. Additionally, we only considered the answers from practitioners with more than 3 years of experience, to ensure our analysis is based on the experience of qualified practitioners. We sent our questionnaire to other employees of MoovAI that did not participate in the interview, personal contacts, and to practitioners identified from social networks. More details regarding these channels can be found in Section \ref{particpants:recruitment}. We also shared the questionnaire with practitioners during an MLOps event organized by our team\footnote{\url{https://semla.quebec/mlops/}}. We shared the questionnaire in April 2023, May 2023, and March 2024 and got 26 answers. We excluded 5 answers to our questionnaire because the participants had less than 3 years of experience, leaving \textbf{21 answers} for further analysis.

\subsection{Analysis Plan} \label{analysis-plan}

Before analyzing any interview, we generated the transcript of the interview using the tool Descript\footnote{\url{https://www.descript.com/transcription}}. It is an automated speech recognition tool that can transform the audio stream of an online meeting into text. We corrected any transcription mistake that obscured or distorted the essence of the message conveyed by the interviewee. Additionally, to preserve the interviewee's privacy, we remove any identifier to other entities (e.g., company names, schools the person attended, etc.). We also send the anonymized transcript to the interviewee to ensure there is no private information left. The anonymized transcripts are available in our replication package.

We use the coding techniques from Straussian Grounded Theory \citep{strauss1994grounded} to extract knowledge from data to answer all our qualitative research questions (i.e., RQ1, RQ3, RQ6). When analyzing an interview, we start with a round of open coding \citep{seaman1999qualitative} to break the interview's content into discrete parts. After the interviews are conducted and we have a set of coded quality issues, we grouped them into categories so that reasoning over the issues is easier. This step is akin to axial coding in Straussian Grounded Theory \citep{strauss1994grounded}. While the current document has been written such as the data collection (i.e., the interviews) and analysis plan are separated, it is important to understand that these two steps happen in an iterative manner. In other words, we analyzed an interview right after we conducted it. Throughout the process, we write down interesting information mentioned during interviews and ideas for preliminary categories in ``memos" \citep{strauss1994grounded}. They are integrated into our theory during the phase of memo sorting \citep{strauss1994grounded}.

In order to identify the most prevalent quality issues and effectively answer RQ2, we compare the number of practitioners who answered that an issue is frequent (i.e. scores of 4 and 5 on a Likert scale) in our questionnaire.

To answer RQ4, we first perform a round of open coding for data quality issues. Each data quality issue mentioned by a practitioner is assigned a code (e.g., ``incorrect labels"). Once the interviews are all conducted, the researchers individually categorized each code into one of the data quality aspects defined in \citet{cappi2021dataset}. We resolved the conflicts through discussion and consensus. We finally answered the question by counting the number of practitioners having experienced an issue with one of the data quality aspects. Note that some data quality issues could not be assigned to a single category (or to any). In this case, a researcher inspected each issue individually and categorized them.

In RQ5, we aim to identify the data types and data collection processes that are the most problematic in practice. We use the data quality issues extracted in RQ4 to find the most problematic data types. Most precisely, a researcher considers each issue and tries to determine to which data type the interviewee was referring. In case of doubt, the extract is ignored. We use the same process to determine the most problematic data collection process. Additionally, to ensure no important information is missed, we tag every issue caused by a data collection process. For example, if an MLSS is unreliable because of the data collection process, we code the extract with the data collection process' name. Similar to RQ1, the codes are grouped into categories through axial coding \citep{strauss1994grounded} so that reasoning over the issues is easier.

To help us code the transcripts, we used Delve qualitative analysis tool\footnote{\url{https://delvetool.com/}}, because the researchers are familiar with it and it is easy to use. Delve is a computer-assisted qualitative data analysis software (CADQAS) that provides simple interfaces to code and analyze data. In order to ensure the quality of the analysis, each document is coded by two researchers. In case of a disagreement in codes, a third researcher plays the role of moderator and selects the final code for a text segment. This process is helpful for the construction of a shared understanding of the data. The coded transcripts of the coders and moderators are available in our replication package.

\section{Study Results} \label{results}
In this section, we present the results we obtained from our interviews. We start by providing demographic information of the interviewees and the practitioners who answered our questionnaire. Then, we present the results for each of our RQs. We provide a structured overview of all quality issues, including their description, causes, consequences, and recommended mitigation strategies in Section \ref{annex:overview}. To refer to a practitioner, we use identifiers. Each identifier has the following structure: ``\practitioner{PXX}", where P is for ``practitioner" and ``XX" are digits. Note that the numbers in the identifiers are not representative of the order we interviewed the practitioners, they are random identifiers assigned to interviewees. When using pronouns, we will always use the feminine form to preserve the practitioner's privacy. We also use identifiers to refer to the quality issues described in Section \ref{RQ1}. These identifiers have the following structure: ``\issue{IXX}", where \enquote{I} is for ``issue" and ``XX" are digits.

\subsection{Demographic information} \label{demographic-information}
In this section, we present the demographic information we collected from the interviews we conducted and the questionnaire.

\subsubsection{Interviews}
\begin{figure}
     \centering    
     \begin{subfigure}[t]{0.45\textwidth}
         \centering
         \includegraphics[width=\textwidth]{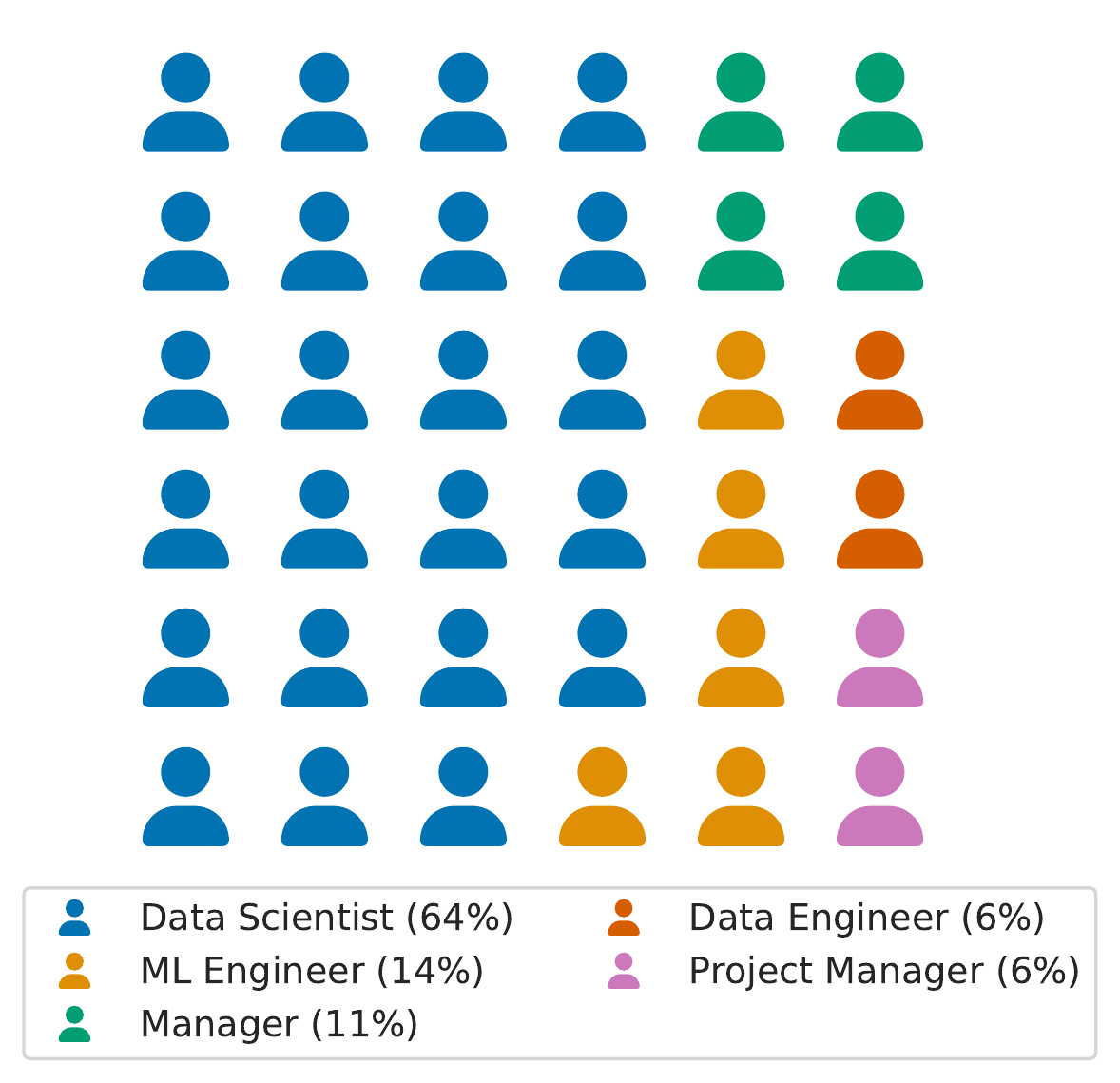}
         \caption{Job Role}
         \label{fig:job_role}
     \end{subfigure}
     \hfill
     \begin{subfigure}[t]{0.45\textwidth}
         \centering
         \includegraphics[width=\textwidth]{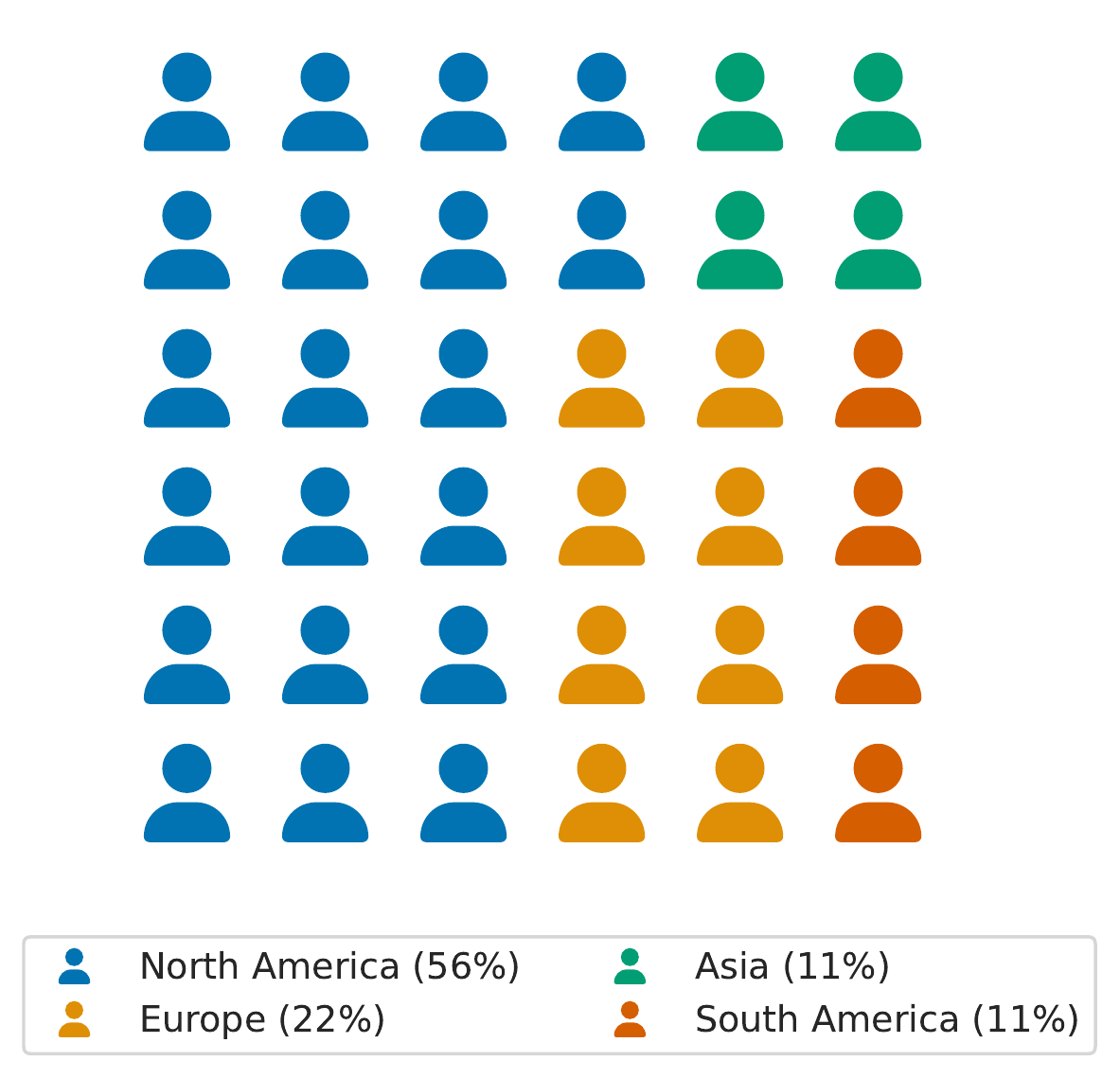}
         \caption{World region}
         \label{fig:world_region}
     \end{subfigure}
     \hfill
     \begin{subfigure}[t]{0.45\textwidth}
         \centering
         \includegraphics[width=\textwidth]{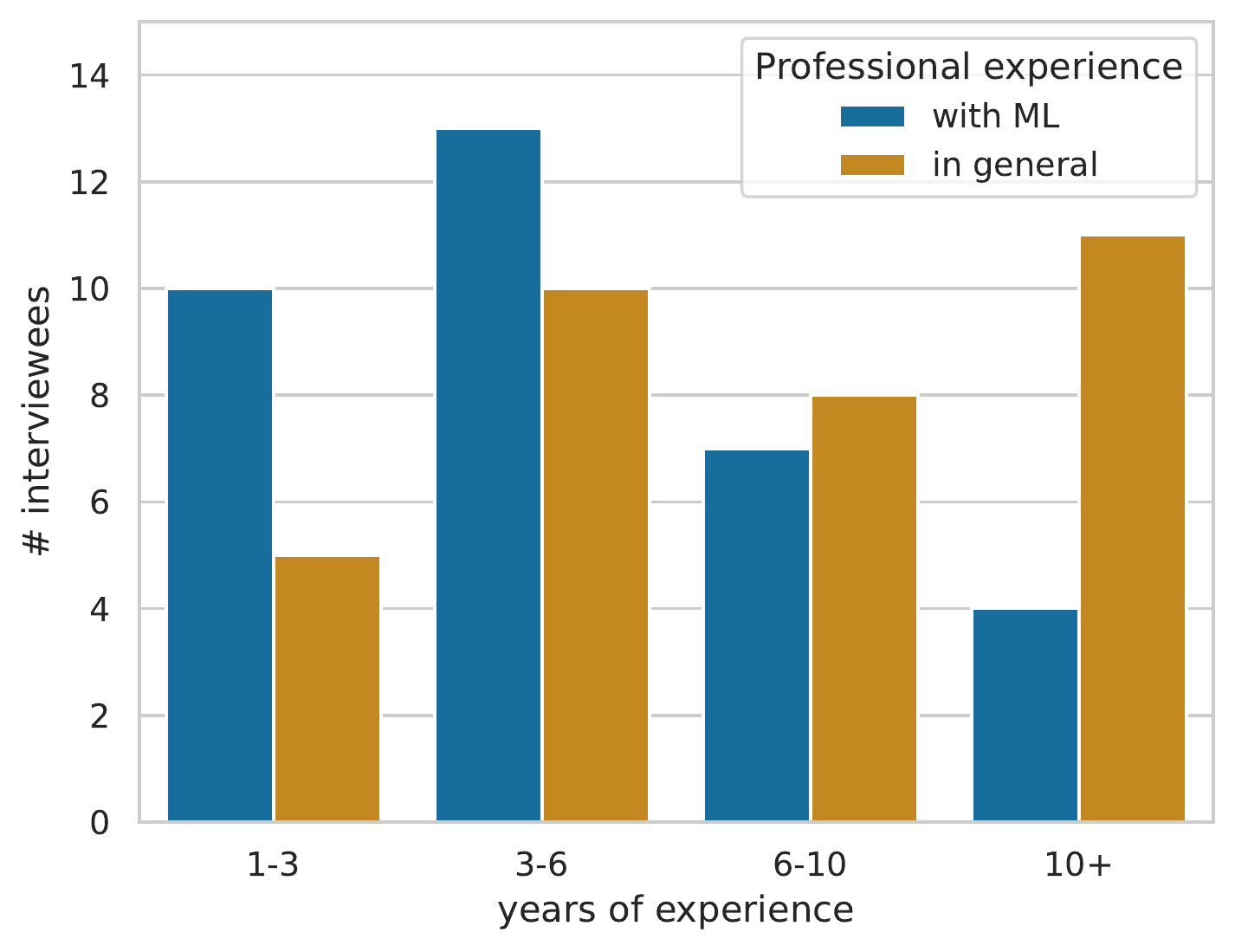}
         \caption{Years of Experience}
         \label{fig:y_of_experience}
     \end{subfigure}
        \caption{Practitioners' Demographics (Interviews)}
        \label{fig:interviewees_demographics}
\end{figure}

\begin{figure}
     \centering
     \begin{subfigure}[b]{0.45\textwidth}
         \centering
         \includegraphics[width=\textwidth]{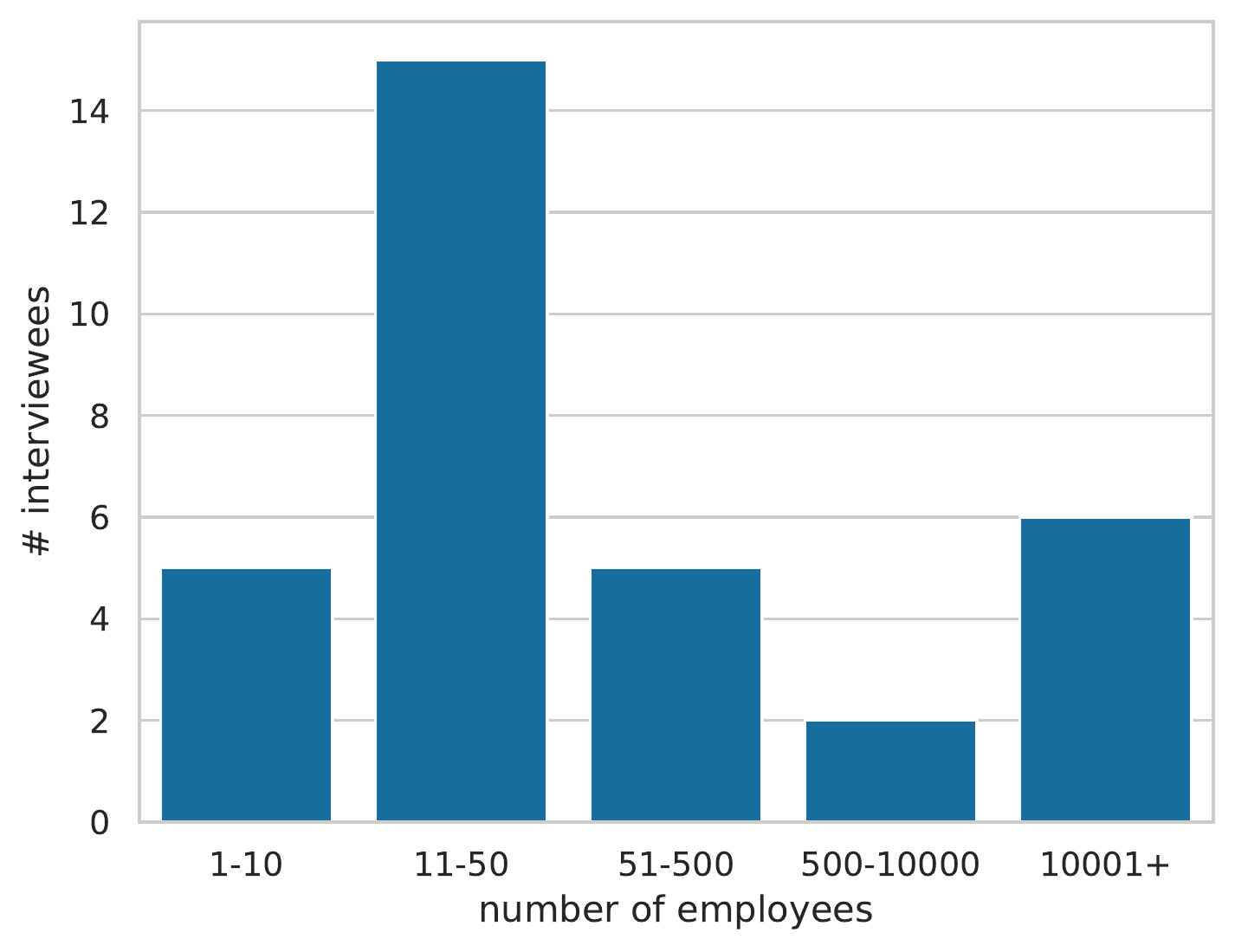}
         \caption{Number of Employees}
         \label{fig:n_employees}
     \end{subfigure}
     \hspace{3em}
     \begin{subfigure}[b]{0.45\textwidth}
         \centering
         \includegraphics[width=\textwidth]{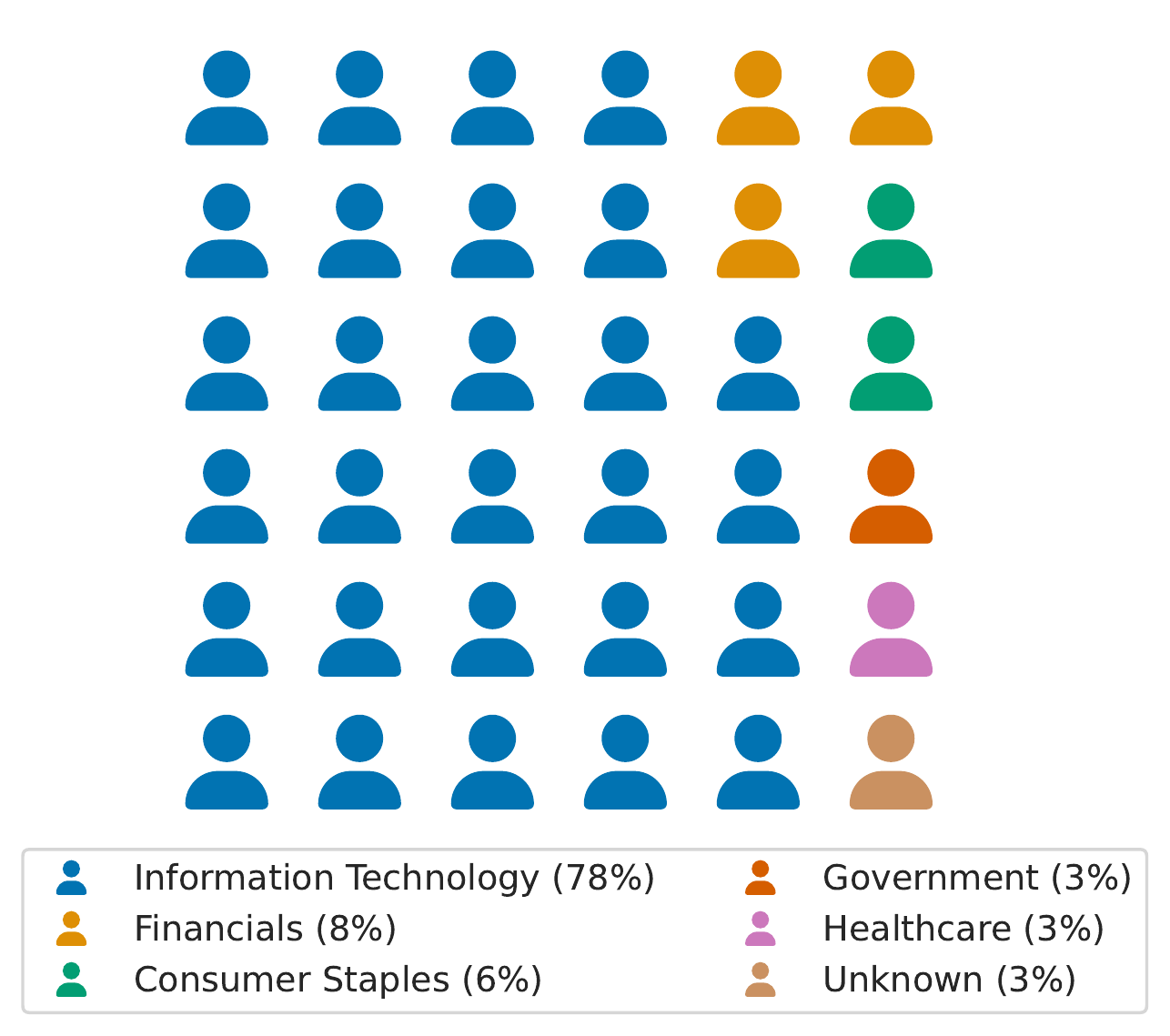}
         \caption{Financial Sectors}
         \label{fig:sectors}
     \end{subfigure}
     \caption{Interviewees' Companies' Demographics}
     \label{fig:companies_demographics}
\end{figure}

% Intro + missing values
Here, we present demographic information from the 36 interviews we conducted that were not excluded from the study. We extracted that information from what the interviewees shared with us before and during the interview. If any information was missing, we searched for the online public profiles of the person (i.e., LinkedIn\footnote{\url{https://www.linkedin.com/}}). If unsuccessful, we tried to contact the person again to answer our questions. In case of failure, this information was left empty and displayed differently depending on the plot type. For waffle plots (i.e., Figures \ref{fig:job_role}, \ref{fig:world_region}, \ref{fig:sectors}), missing values were replaced with the ``unknown" tag and displayed on the figure. For histograms (i.e., Figures \ref{fig:y_of_experience}, \ref{fig:n_employees}), missing values are not shown for clarity. The number of missing values for Figure \ref{fig:y_of_experience} and Figure \ref{fig:n_employees} is 2 and 3, respectively.

% Intro to presenting demographics
The demographic information is presented with two sets of figures, i.e., Figure \ref{fig:interviewees_demographics} and Figure \ref{fig:companies_demographics}. The former displays the demographic information of the practitioners, while the latter displays the demographic information of the company the practitioners currently work at. 

% Years of experience
In Figure \ref{fig:y_of_experience}, we display our interviewee’s professional experience. We indicate their professional experience in general and their experience with ML. For example, an interviewee who started his career 7 years ago and transitioned into an ML role 4 years ago would be in the bin “6-10” for his experience “in general” and would be in the bin “3-6” for his experience “with ML”. As we can see from the figure, we interviewed practitioners with varying years of professional experience. The practitioner most generally had between 3 and 6 years of experience in ML. While 10 interviewees had less than three years of experience with ML (1-3 years), the majority (6/10) transitioned from a role similar to their current ML role (e.g. from data analyst to data scientist).
% Job roles
We display the practitioners' job roles in Figure \ref{fig:job_role}. We used the title they described themselves. We included in ``manager" all roles that orbited around management, such as CTO. As we can see, the majority of the interviewees (64\%) are data scientists. In contrast, 20\% are engineers (i.e., ML engineers or data engineers) and 17 \% have management roles (i.e., manager or project manager).
% World regions
Figure \ref{fig:world_region} displays where the practitioners are based. Because we interviewed practitioners from MoovAI, the majority of interviewees are from North America (56\%). Excluding the 14 practitioners we interviewed from MoovAI, we have a more balanced set: 22 \% are from Europe, 5\% from North America (excluding MoovAI's practitioners), 11\% from Asia, 11\% from South America, and 3\% from the Middle East.

% Company's demographics 
In Figure \ref{fig:companies_demographics}, we display demographic information on the company the practitioners currently work at. 
% Number of employees
In Figure \ref{fig:n_employees}, we demonstrate the number of employees. Once again, a category (11-50) has a larger count because of MoovAI's employees. Apart from that, we managed to interview practitioners from a wide range of company sizes: 5 practitioners work in startups (1-10 employees), 20 in small to medium-sized companies (11-50 and 51-500), and 7 in large companies (500-10000 and 10001+). 
% Financial sector
We display the companies' industry sectors in Figure \ref{fig:sectors}. We used the Global Industry Classification Standard\footnote{\url{https://en.wikipedia.org/wiki/Global_Industry_Classification_Standard}} to classify companies. We added the category ``government" for practitioners working for the government. The majority of our interviewees (78\%) work for a company in the information technology sector: 8 \% work in the financial sector, 6 \% in the consumer staples sector, 1 person in healthcare and another for a government.

\subsubsection{Questionnaire}
In this section, we present and discuss the demographic information collected using the 21 answers to the questionnaire. In comparison to the interviews, we only collected the job roles and years of experience of the practitioners. We wanted to avoid collecting too much information to avoid deterring practitioners from answering the questionnaire. Thus, we only collected data that did not allow us to identify the practitioner or his company.

\begin{figure}
     \centering    
     \begin{subfigure}[t]{0.45\textwidth}
         \centering
         \includegraphics[width=\textwidth]{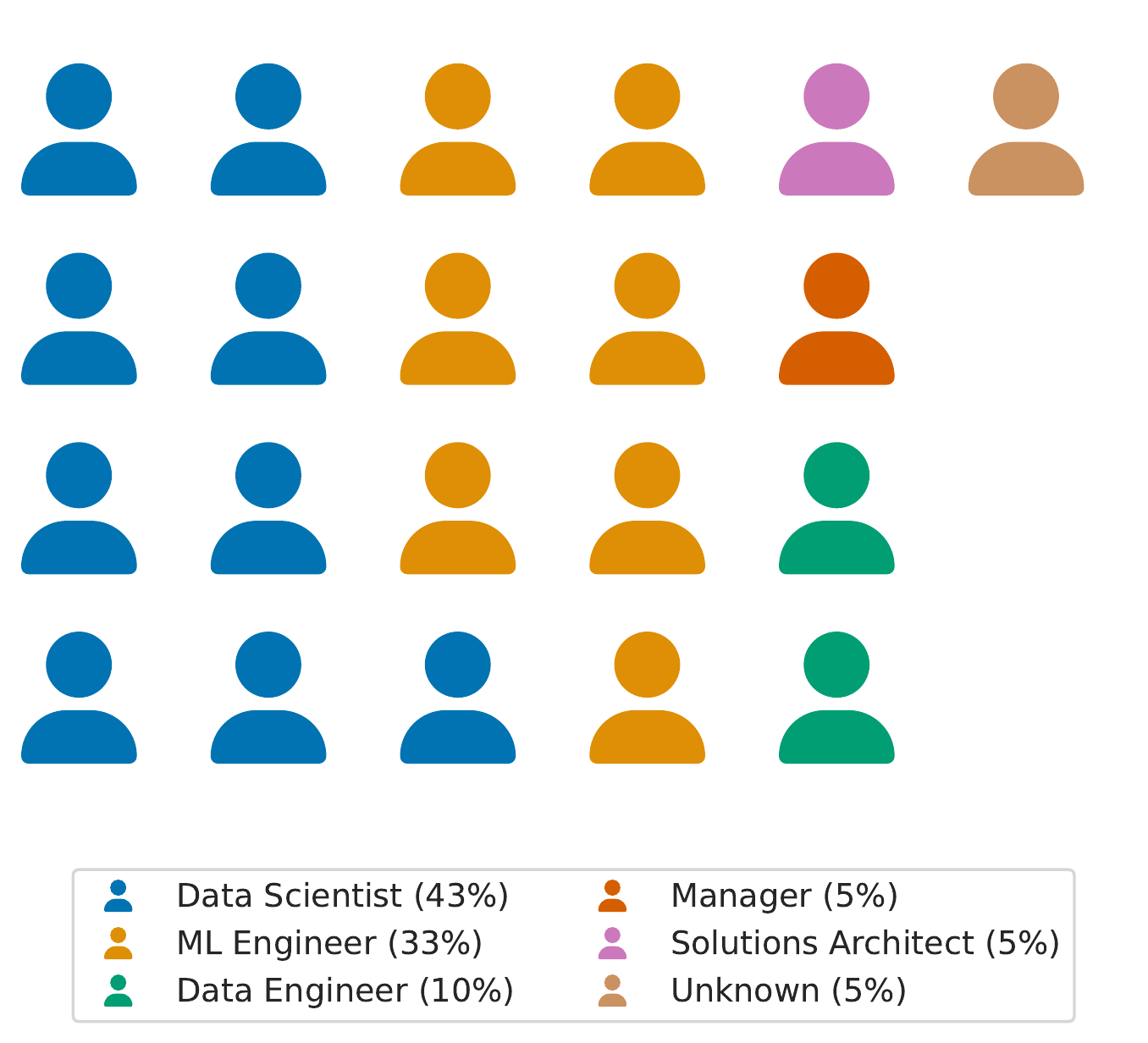}
         \caption{Job Role}
         \label{fig:job_role_questionnaire}
     \end{subfigure}
     \hfill
     \begin{subfigure}[t]{0.45\textwidth}
         \centering
         \includegraphics[width=\textwidth]{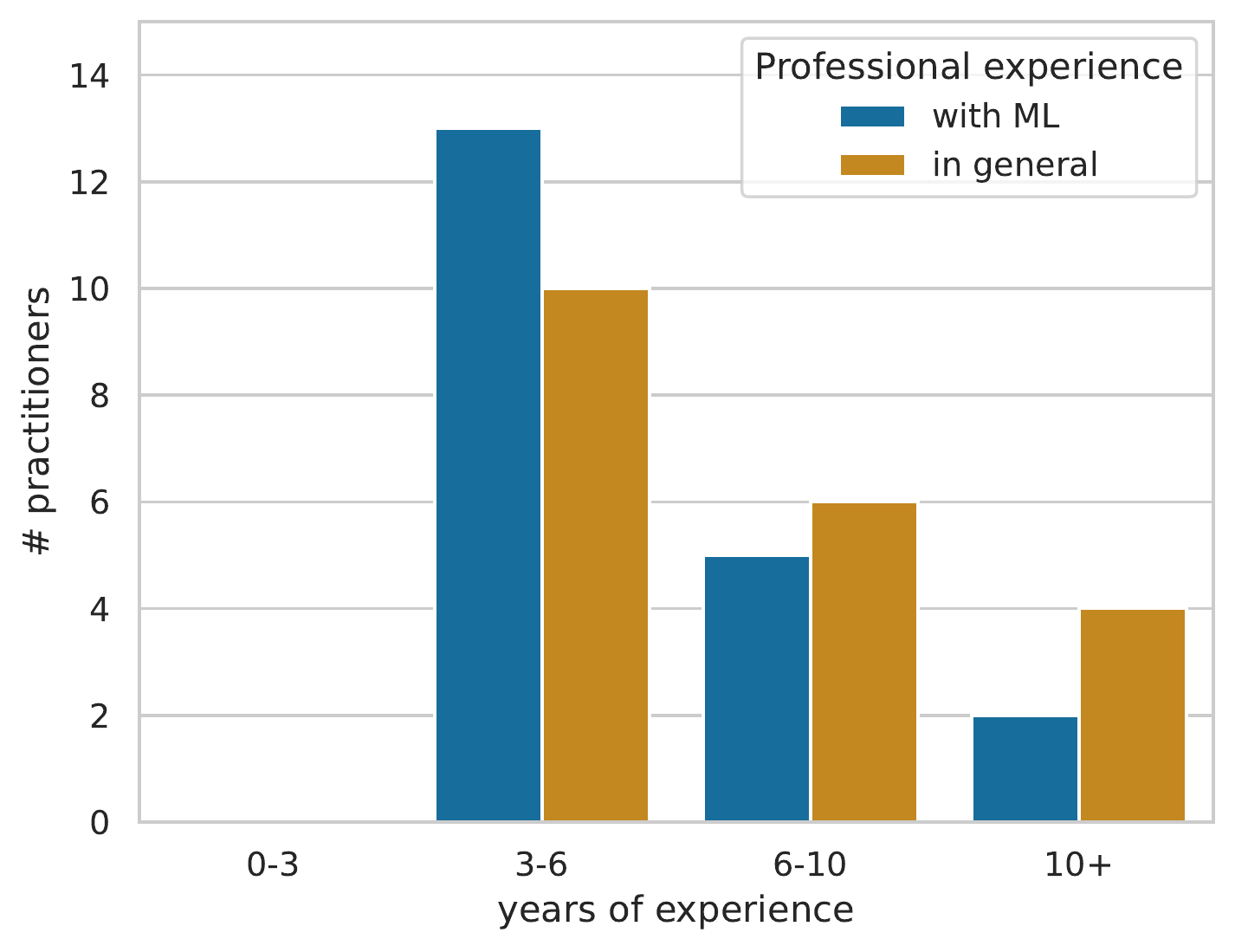}
         \caption{Years of Experience}
         \label{fig:y_of_experience_questionnaire}
     \end{subfigure}
        \caption{Practitioners' Demographics (Questionnaire)}
        \label{fig:interviewees_demographics_questionnaire}
\end{figure}

In Figure \ref{fig:y_of_experience_questionnaire}, we display the professional experience of the practitioners who answered the questionnaire. Similar to the interviews, the interviewees most generally had 3-6 years of experience. Contrary to the interviews, no practitioners with less than 3 years of experience were included, as mentioned in Section \ref{participants:inclusion-exclusion}.  In Figure \ref{fig:job_role_questionnaire}, we present the job roles of the practitioners who answered our questionnaire. We obtained a diverse sample of professionals. Similarly to the interview, the most common job role for the questionnaire’s respondents is Data Scientist (9 answers). A similar number of ML Engineers answered our questionnaire (7 answers). Other job roles include Data Engineer (2 answers), Manager (1 answer), and Solution Architect (1 answer)

\subsection{RQ1: What are the quality issues encountered by practitioners when building MLSSs?} \label{RQ1}
In this section, we describe the quality issues extracted from our interviews. For each issue, we present its causes and consequences according to the experience of our interviewees. For clarity, we grouped the quality issues into quality dimensions. For example, the issues making the maintenance of an MLSS more difficult are presented under the ``Maintainability" subsection (Section \ref{RQ1:maintainability}). We provide the full list of quality issues in Table \ref{table:quality_issues}. The issues' identifiers defined in this table will be used throughout the rest of the paper to refer to an issue without having to describe it all over again.

% ----------------------------------------------------------------
%                      Quality issues table
% ----------------------------------------------------------------
% \setlength{\tabcolsep}{8pt} % horizontal spacing, Default value: 6pt
\renewcommand{\arraystretch}{1.5} % vertical spacing, Default value: 1
 \begin{table}
 \caption{List of quality issues in MLSSs}
 \centering
 \begin{tabular}{|>{\centering}m{1.3em}|>{\centering}m{2em}|m{30em}|}
 \hline
 & Id & \multicolumn{1}{c|}{Description}   \\
 \hline
 \multirow{3}{*}[-4em]{\rotatebox[origin=c]{90}{ Evaluability}}
        & 1.1  & Evaluating the quality of a model offline (i.e., not in a production environment) is inaccurate even when the dataset used for evaluation is representative of the data distribution in production.  \\
        & 1.2  & Defining a good business metric for evaluating an MLSS is difficult. For MLSS, a business metric refers to the degree to which the MLSS successfully achieves the goal it has been built for.    \\
        & 1.3  & Trying to simulate the environment/system in which the model will operate (to evaluate the model in this simulated environment) is difficult and error-prone. \\
        & 1.4  & Evaluating the quality of a dataset (e.g., presence of incorrect labels, noisy/incorrect features,  wrong format data, etc.) is difficult and time-consuming. \\
    \hline
  \multirow{2}{*}[-1em]{\rotatebox[origin=c]{90}{\parbox[c]{1cm}{\centering Explaina-bility}}}
        & 2.1  & Explaining a model's predictions to people without ML knowledge (e.g. business stakeholders, users) using explainability techniques is challenging because these techniques require technical knowledge and interpretation.  \\
        & 2.2  & The explanation techniques sometimes provide explanations that do not make sense and can not be relied on. \\ 
    \hline
 \multirow{3}{*}[-1.8em]{\rotatebox[origin=c]{90}{Debuggability}}
        & 3.1  & Reproducing bugs in an MLSS is difficult because of unstable data sources. A data source is unstable if it returns different values for the same queried record.  \\  
        & 3.2  & Debugging data streaming systems (e.g., Hadoop) is difficult because it is difficult to picture what data should look like at each step of the data pipeline.  \\
        & 3.3  & Debugging an MLSS is time-consuming when its data sources are managed by other teams, because it may require inspecting these data sources.   \\ 
    \hline
 \multirow{3}{*}[-0.5em]{\rotatebox[origin=c]{90}{Efficiency}}
         & 4.1  & Training models consume too many resources (e.g., time, computing power, etc.). \\
        & 4.2  & The queries sent to an MLSS are not answered timely (i.e., latency/delay issues).   \\
        & 4.3  & At inference time, models consume too much memory.  \\
     \hline
 \multirow{3}{*}[-.4em]{\rotatebox[origin=c]{90}{Maintainability}}   
        & 5.1  &  Maintaining an MLSS is difficult because there is not enough information describing the data the MLSS consumes. \\
        & 5.2  &  Maintaining a model is difficult because there is not enough information describing how the model was generated. \\
        & 5.3  & Managing the dependencies (i.e., software libraries) of an MLSS is challenging and error-prone. \\ 
    \hline
 \multirow{3}{*}[-.8em]{\rotatebox[origin=c]{90}{ Reliability}}   
        & 6.1  &  Having a reliable model is difficult because of concept or data drift. \\
        & 6.2  & Having a reliable model is difficult because of external data providers. \\
        & 6.3  & Having a reliable MLSS is difficult because of the data pipelines which are brittle and have technical debt.  \\ 
    \hline                    
\end{tabular}
\label{table:quality_issues}
 \end{table}

% ------------------------------------------------------------------
%                           Evaluability
% ------------------------------------------------------------------
\subsubsection{Evaluability} \label{RQ1:evaluability}
Practitioners often mentioned that evaluating the quality of their models was difficult, especially offline. As \practitioner{P17} said: \citepr{There is no good enough indication to say that a model will perform as expected in the real world. [...] Before your deploy model, you cannot really know how it will perform, but at least you can guess. And guessing is not good enough}. Similarly, evaluating the data quality of a dataset is difficult and can be time-consuming. So, we define evaluability as the ability to evaluate the quality of ML models and datasets. Generally, when evaluating models, practitioners follow one of two evaluation procedures. We describe the evaluation procedures along with their issues below.

% Evaluation procedure 1
In the research community, the most common way to evaluate the quality of a model is to measure its performance on a test dataset using a metric such as accuracy or F1-score (depending on the application). This is also how most practitioners learned to evaluate their models and, as a result, how they evaluate their models in practice. However, practitioners commonly shared that they were often disappointed with the performance of their model when deployed in a production environment \practitioner{P7, P9, P16, P17, P23}. We observed two reasons that can explain why this procedure to evaluate datasets is flawed. First, the goal that is targeted when an ML metric such as accuracy is used is different from what is important for the end application. As a result, a model that is optimized for an ML metric may not answer optimally to the needs of the application (in terms of quality of predictions) \practitioner{P3, P7, P9}. For example, for a recommendation engine, improving accuracy might not lead to increased revenues (i.e., the business need), if the recommended products only provide minor profit \practitioner{P9}. Perhaps more profitable products should have been recommended instead. Second, the evaluation procedures do not take into consideration the system in which the model is embedded \practitioner{P1, P16}. An improvement to a model in terms of accuracy on a testing dataset might not be of relevance since it does not take into consideration how the other components interact with the predictions of the model. For example, improving the component responsible for query understanding in a search engine might not improve the engine's search results overall since other components, such as the item understanding component, also influence the overall engine's performance \practitioner{P16}.

% Evaluation procedure 2
To address these shortcomings in the evaluation procedure, some practitioners estimate the quality of a model by simulating the environment in which it will be deployed and measuring business metrics. Business metrics are quantifiable measures that can be used to track the performance level of a business \citep{luther}. For MLSSs, business metrics give a measure of how well an MLSS achieves the goal it has been built for. For example, for an ML trading system, a business metric could be the profit made. We observed two issues making evaluating a model in a sandbox using a business metric difficult. First, reproducing the production environment is error-prone \practitioner{P1, P7}. For example, \practitioner{P7} did not take into account that the data used for prediction was available only after a delay and their model had to predict further in the future than expected. After reflecting on her experience with building a sandbox for a project, \practitioner{P1} said that she thinks moving to A/B testing sooner (i.e., online evaluation) would have been preferable instead of improving the sandbox. The reason is that A/B testing is easier to implement and provides measures of quality that are more accurate. Second, defining good business metrics to prove the impact of a model is not an easy task for some applications, such as recommendation engines \practitioner{P17, P25}. As \practitioner{P17} said: \citepr{There is no strong definition of what's the best recommendation, right? The only thing we can do is monitor how much people return. [...] So when people are happy with their purchase, they return less. So that's kind of a belief, just a belief there}.

% Consequences 
As a result of not being able to accurately measure the quality of a model, practitioners have ended up (1) spending a lot of time on improvements that do not provide any real value \practitioner{P9}, (2) deploying models with low quality \practitioner{P1, P7, P9, P13, P16, P17, P23}, and (3) being unable to prove the value added by an MLSS to business stakeholders \practitioner{P25}. Hence, stakeholders may lose confidence that the MLSS brings value to their company. As \practitioner{P25} said: \citepr{[My manager] is skeptical if these are the right metrics [...] [As a result] you are trying to prove [that your system brings value], but your proof is not considered or taken forward}.

% Dataset evaluability 
In the early phases of building an MLSS, practitioners usually spend time understanding datasets and evaluating their data quality. Ensuring the quality of a dataset is especially important when it has been created by a third party. For example, \practitioner{P23} mentioned that a public ML dataset contained irrelevant and harmful images by mistake. Practitioners have reported struggling to evaluate the data quality of a dataset \practitioner{P8, P23, P30} and spending a lot of time doing so \practitioner{P8, P30}. When asked about the most pressing issue in ML in her opinion, \practitioner{P30} answered the following: \citepr{We should find some method [to] understand faster the quality of the data. [...] I mean, just push one button and understand the data quality}.  Estimating the quality of a dataset can be used to (1) assess the feasibility of applying ML to a problem, (2) estimate the effort needed to clean the dataset, and (3) select datasets. We posit that evaluating the quality of a dataset can take a lot of time because it requires a deep understanding of the domain and datasets are not always sufficiently documented (more details in Section \ref{RQ1:maintainability}). This problem is exacerbated by the size of datasets that can be quite large \practitioner{P8}. \practitioner{P8} reported using tools to help her estimate data quality but ended up giving up because it required too much effort: \citepr{yes, there are tools, but most that I've tried, always require so much time investment that they become less worth than just discarding data that we find is suspicious}.

% ------------------------------------------------------------------
%                           Explainability
% ------------------------------------------------------------------
\subsubsection{Explainability} \label{RQ1:explainability}
Explainability refers to any technique that tries to explain the predictions of a model \citep{DBLP:journals/corr/abs-1909-06342}. 
% The explanations are too technical for users
While one of the goals of explainability is to provide intelligible explanations for the predictions of an ML model \citep{DBLP:journals/corr/abs-1909-06342}, practitioners have reported that using explainability tools to explain the predictions of a model to people without ML knowledge (e.g., users and business stakeholders) is challenging because these tools require interpretation and technical knowledge \practitioner{P3, P9, P27, P32}. As \practitioner{P9} said: \citepr{the interpretation of the outputs of the model [is] very hard to vulgarize to the business and to the stakeholders}. On the topic of explaining predictions for business stakeholders, \practitioner{P3} reported: \citepr{most of the time, [explanations] need to be in the lamest terms. If it's mathematical gibberish, it won't mean anything to them. So the challenge is also to make it intuitive and useful for them}. As a result, explanation techniques are seldom used to increase users' confidence in the MLSS. Over the 42 interviews conducted, only practitioner \practitioner{P32} reported using explainability techniques in production to increase trustworthiness. 

% The explanations are unreliable
Practitioners have also reported being distrustful of the outputs of explainability tools because the explanations do not always make sense \practitioner{P4, P10, P11}. Regarding one of her previous experiences with Shapley values, \practitioner{P4} shared the following: \citepr{[I] see [that the] contribution is positive [for] this variable, but it should be negative. What's happening? And then it produces more questions than answers. [It] is really a difficult task to have good explanations}. Our practitioners mentioned observing that the tools' explanations get seemingly worse as the number of features used by the model increases \practitioner{P10} or when the complexity of the model (in terms of the number of parameters) increases \practitioner{P4}. To show her mistrust is justified, \practitioner{P11} referred us to the study made by \citet{merrick2020explanation}. Notably, it shows cases when Shapley-value-based feature attribution methods produce faulty explanations. Consequently, practitioners become distrustful of the explanations given by these tools and avoid using them in high-stakes scenarios. For example, \practitioner{P20} preferred using interpretable models instead of explainability techniques for an application facing regulatory constraints.

% ------------------------------------------------------------------
%                           Debuggability
% ------------------------------------------------------------------
\subsubsection{Debuggability} \label{RQ1:debuggability}
Debugging refers to identifying and removing bugs from a computer's hardware or software \citep{OL}. Hence, we define debuggability as the degree to which a system can be easily debugged. 

% Reproducing a bug
In order to fix a bug, developers will generally try to reproduce it to identify its root cause. This process is difficult with MLSSs because of unstable data sources \practitioner{P1, P3, P16}. To reproduce a bug, the same input data must be fed to the system. However, the data sources may return different values for the same queried object, because of unstable data sources. For example, \practitioner{P16} worked on a recommendation system that ingested embeddings. These embeddings were generated daily; the new version replaced the old one. By the time \practitioner{P16} was aware of a bug, the embeddings that triggered the bug were already replaced. Thus, reproducing a bug proved to be impossible at times. As a result, bugs may tend to be ignored, reducing the trust of stakeholders in the system. Similarly, data sources that are unstable because of historical revisions of data will render the MLSS less debuggable \practitioner{P1, P3}. We refer the reader to Section \ref{RQ5} for further information on the challenges with unstable data sources.

% Finding the bug
Once the data required to reproduce the bug is obtained, the process of searching for the root cause of a bug can begin. Our practitioners have reported that this process can be difficult for two reasons. First, debugging data streaming systems is more difficult than debugging other software systems according to our practitioners' experience \practitioner{P8, P19, P25}. When reproducing a bug in order to find the root cause, one might verify the state of data at key steps in the data pipeline. However, knowing the expected state of data at these key steps is difficult. As a solution to this problem, one could compare the data against what data normally looks like after the transformations. For example, one could compare a value against the mean or the mode. However, this is not possible with our practitioners' data streaming systems, since only one entry is processed at a time. Hence, data streaming systems are recognized to be challenging for debugging \practitioner{P8, P25}. Second, debugging MLSSs can be difficult because communication across teams is often necessary to find the root cause of an issue \practitioner{P27}. As explained in Section \ref{RQ5}, MLSSs are highly coupled with the data they are consuming, and any issue with its quality might affect the MLSS's quality. Hence, in case of an issue with the MLSS, practitioners commonly have to inspect the MLSS's data sources, which are seldom managed by their team. Communicating with the team managing the other data source may be a long back-and-forth process because the other team has its own agenda and does not necessarily prioritize the bug. Additionally, the practitioners may have to reach out to more than one team, extending the process. This problem was also observed by \citet{mailachsocio}.

% ------------------------------------------------------------------
%                           Efficiency
% ------------------------------------------------------------------
\subsubsection{Efficiency} \label{RQ1:efficiency}
The ISO/IEC 25010:2011 standard defines efficiency as: ``[the] performance relative to the amount of resources used under stated conditions". Efficiency issues in MLSSs are critical because of the large volume of data MLSSs have to process. According to our interviews, efficiency has become a growing concern, as running MLSSs is expensive \practitioner{P8, P17, P23, P27, P26, P29, P30, P37}. Efficiency issues arise both at model training and inference time. 

Practitioners have reported long training times, which leads to hefty spending \practitioner{P8, P23, P30} and slow product development \practitioner{P3}. At inference time, practitioners mentioned concerns with the latency of data pre-processing pipelines \practitioner{P16, P38} and the memory consumption of the model \practitioner{P31, P38}. Memory consumption can be an issue for applications that run on smaller devices, such as cellphones \practitioner{P35}. \practitioner{P38} observed that most latency issues come from the data pre-processing pipeline and not from the model.

% ------------------------------------------------------------------
%                           Maintainability
% ------------------------------------------------------------------
\subsubsection{Maintainability} \label{RQ1:maintainability}

The ISO/IEC 25010:2011 standard defines maintainability as: ``[the] degree of effectiveness and efficiency with which a product or system can be modified by the intended maintainers". Practitioners have mentioned having issues with maintaining datasets, models, and libraries.  

As it is explained in Section \ref{RQ5}, the quality of ML models highly depends on the quality of the data it processes. Hence, maintaining an MLSS requires not only a good understanding of the software and the ML models but also of the data. This understanding can be communicated verbally but is more effectively shared through documentation. As practitioner \practitioner{P29} expressed: \citepr{It's a time-consuming process to go back and forth and ask about the correct description of the columns, and sometimes there is no clear answer}. We have observed two types of documentation issues with datasets. First, there often is a lack of documentation describing the data in the dataset \practitioner{P1, P2, P5, P7, P8, P19, P24, P29, P38}. As a result, practitioners may be unable to understand what a feature represents \practitioner{P1, P5, P29}, how it is measured (i.e., what are the units) \practitioner{P1} and what is the meaning of special tokens \practitioner{P2}. As an example of unspecified special tokens, a ``none" value may mean that there is nothing for that attribute or that there has been an oversight to provide the real value \practitioner{P2}. This is a problem when building MLSSs, but also when maintaining them, since modifying the system without data documentation is slower and more error-prone.

Second, practitioner \practitioner{P8} expressed not being able to assess how reliable a data source is. As the interviewee shared: \citepr{So we had [...] over 50 different sources on that project. We had a lot of issues tracking, which data came from who and which data was, let's say, uh, production-grade, or, uh, just experimental}. As detailed in Section \ref{RQ5}, knowing the reliability of a data source is important during the maintenance of a system to ensure the reliability of the system.

We observed other issues that we posit are symptoms of poor data management, such as temporal data leakage (i.e., evaluating a model with data that would not be available if the model was deployed) \practitioner{P20, P23} or corruption of production data with dummy data \practitioner{P8, P35}. Dummy data can be mixed with production data when the production environment of an MLSS is used for testing and the dummy users created for testing are not deleted after usage. These dummy users can degrade the production model's performance.

% Model issues
Similar to the datasets, practitioners have reported having issues with model maintenance. There can be poor experiment tracking \practitioner{P20, P23, P25}. Experiment tracking provides the documentation necessary to understand how a model was created. As a consequence of a lack of such documentation, \practitioner{P20} was unable to maintain a model, and practitioners \practitioner{P20, P25} found it difficult to explain to stakeholders why a model had performance issues. As a result, stakeholders may lack trust in the system \practitioner{P25}. Experiment tracking becomes of paramount importance when ML projects are handed off to other teams. Failure to do so may force the maintainers to rebuild the model from scratch since there is not enough information to retrain the models while following the same procedure, as practitioner \practitioner{P20} experienced.

% Dependencies
Dependency management has also been an issue for some practitioners \practitioner{P10, P16, P26, P27, P38}. As expressed by practitioner \practitioner{P38}: \citepr{I think a big problem is, dependency hell [...] it is a big problem because [environments are easily] broken}.  Poor dependency management can break an MLSS in a production environment \practitioner{P16, P26, P27,  P38}. This situation may happen if the development environment used by the practitioners is not the same as the production environment \practitioner{P27} or if dependencies are not pinned and automatically upgraded by the package manager \practitioner{P16, P26, P38}. Forgetting to pin library versions can become taxing when popular open-source libraries such as ``Transformers" are used since they have a lot of version upgrades \practitioner{P16}.

% ------------------------------------------------------------------
%                           Reliability
% ------------------------------------------------------------------
\subsubsection{Reliability} \label{RQ1:reliability}
The ISO/IEC 25010:2011 standard defines reliability as: ``[the] degree to which a system, product or component performs specific functions under specified conditions for a specified period of time". Building reliable MLSSs is challenging because ML models are tightly coupled to the data they have been trained on. Any change in the distribution of data might hinder the predictive capabilities of a model. Thus, in addition to the reliability issues that may be encountered in software systems without ML, owners of MLSSs must tackle the challenges of data quality assurance during model evolution. Reliability issues with data may stem from the data providers or the data transformation pipelines.

% Issues caused by unstable data sources
We observed two issues with data sources that affected the reliability of MLSSs: (1) data and concept drift, and (2) unstable external data providers. We describe these issues in Section \ref{RQ5}. As a result of unstable data sources, practitioners reported their MLSSs experienced poor ML performance \practitioner{P4, P7, P10, P16, P21, P23, P25, P37} or were unable to deliver predictions \practitioner{P8, P10}. Managing unreliable data sources is perceived as challenging by practitioners and, if poorly done, can lead to the failure of ML projects, as experienced by \practitioner{P7}: \citepr{If you are taking data from data sources and one of them is missing, [...] the best you can do is sending alerts and stuff like that. But if your system has too many alerts, [the clients] may just shut it down and say, whatever, let's forget about ML}.

% Issues caused by unstable data transformation
In addition to data sources, data transformation pipelines can be the cause of reliability issues. A brittle data pipeline that incurs technical debt \citep{NIPS2015_86df7dcf} may be a cause of reliability issues for MLSSs. A poorly designed pipeline may not handle well data with unexpected characteristics (e.g., wrong data type, missing values, outlier values) and create feature vectors that can not be recognized by the model. As a result, an MLSS may produce poor predictions \practitioner{P13, P16, P25} or even crash \practitioner{P8}.

\subsection{RQ2: Which quality issues are the most prevalent?} \label{RQ2}
\begin{figure}[t]
 \centering
 \includegraphics[width=1\textwidth]{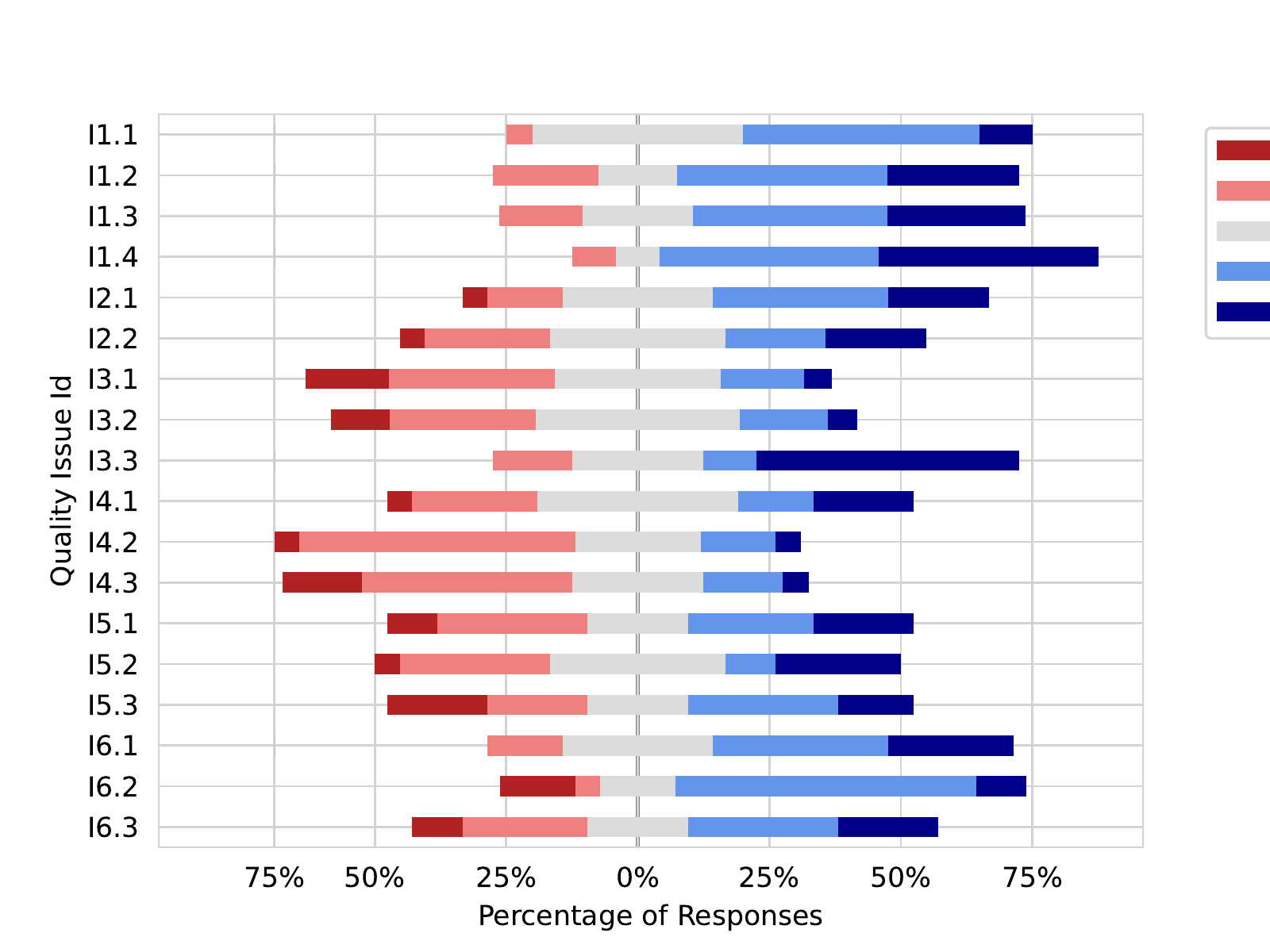}
 \caption{ How common the issues are according to the practitioners' experience.}
 \label{fig:likert_plot}
\end{figure}

\begin{figure}[t]
 \centering
 \includegraphics[width=0.7\textwidth]{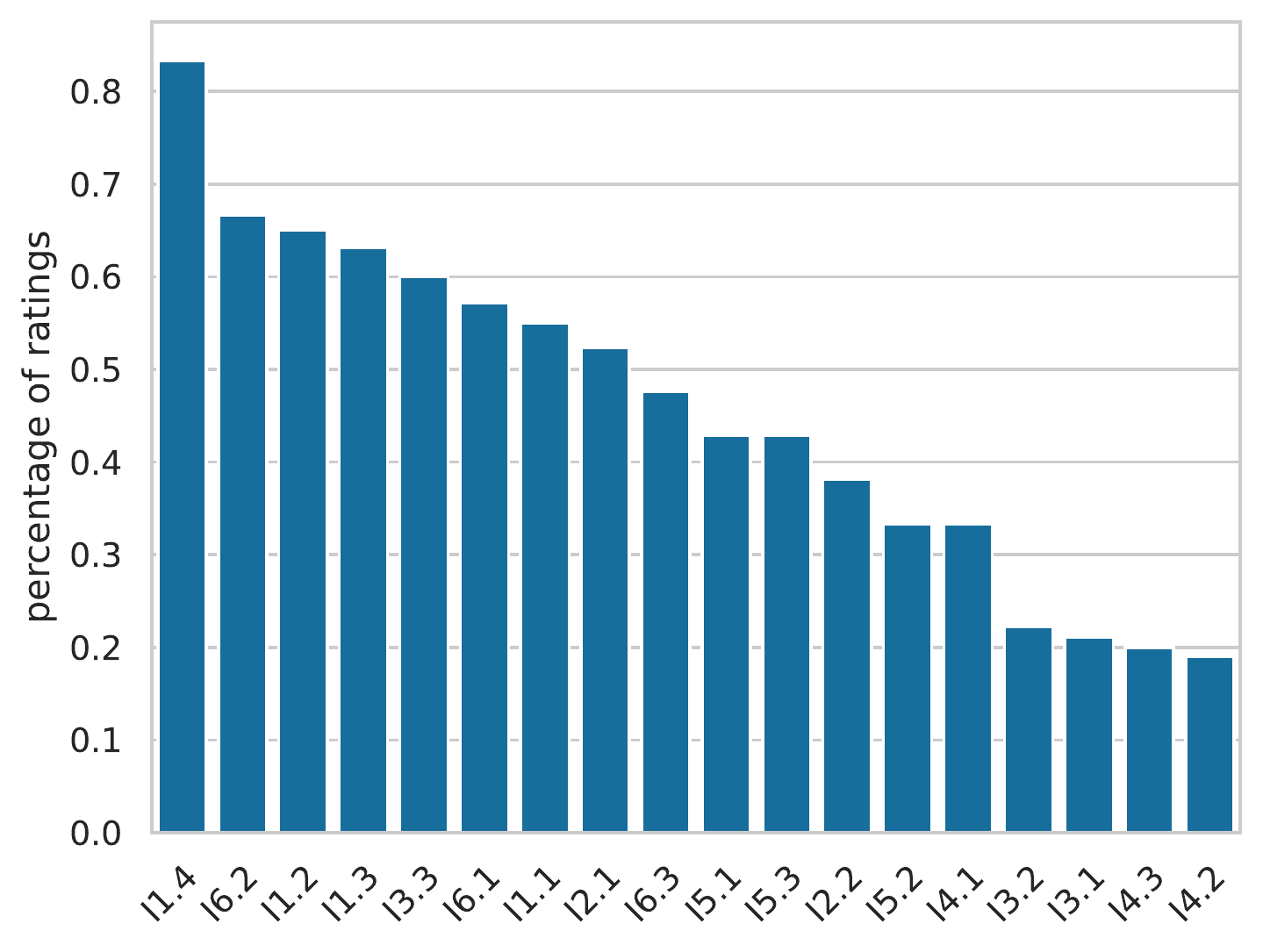}
 \caption{Percentage of answers rating an issue to be frequent}
 \label{fig:percentage_frequent}
\end{figure}

As explained in Section \ref{analysis-plan}, we answered this RQ using a questionnaire. In the questionnaire, we asked practitioners to rate on a scale from 1 to 5 how frequently a quality issue is according to their experience. A practitioner answers 1 if he never experienced an issue and 5 if he frequently has it. The responses are not mandatory; hence, if a practitioner does not understand a question or feels he does not have the necessary experience to answer one, he could skip it. We present the results using Figure \ref{fig:likert_plot} and Figure \ref{fig:percentage_frequent}. For conciseness, we use the issues' identifiers (e.g., \issue{I3.1}) instead of their full description to refer to the issues. The raw results of our questionnaire (e.g., the number of ratings of 1 for an issue or the number of times a question was skipped) are given in Annex \ref{annex:questionnaire}. As mentioned earlier, 21 practitioners answered our questionnaire.

Figure \ref{fig:likert_plot} displays the distribution of the ratings (from 1 to 5) for each quality issue. The color code used in the figure maps to the scale of 1 to 5 used in the questionnaire. One can observe that some issues are more prevalent than others. For simplicity, we consider that the issues that are the most frequent are the ones crossing the 50\% threshold on the right side of the figure,  while the ones that are rarer are under the 50\% threshold. The issues in the former group are issues \issue{I1.1, I1.2, I1.3, I1.4, I2.1, I2.2, I3.3, I4.1, I5.1, I5.3, I6.1, I6.2, I6.3}, while the ones in the later are \issue{I3.1, I3.2, I4.2, I4.3, I5.2}, with \issue{I5.2} exactly on the threshold. We identify two potential factors explaining why some issues are more frequent than others. First, there are issues that are common to a lot of ML projects, while others, only to projects with specific constraints. For example, the issues \issue{I4.2, I4.3} (which refer to latency and memory issues respectively) might only be experienced for projects with speed or resource constraints. Other issues, such as \issue{I1.4} (which refers to the challenging aspect of evaluating the quality of datasets), are applicable to almost any ML project. Second, some issues already have solutions or mitigation techniques designed to attenuate their consequences. For example, memory consumption issues can be attenuated using quantization methods \citep{gholami2021survey}, as we suggest in Section \ref{discussion:efficiency}. As a result, practitioners who are faced with such issues will not consider it to be a problem, since a mitigation technique already exists.

As mentioned in our methodology (Section \ref{analysis-plan}), we count the percentage of answers to our questionnaire rating an issue to be frequent to compare the prevalence of quality issues. We consider an issue to be frequent according to a practitioner if he/she gives it a score of 4 or 5 on the Likert scale.  We display the results in Figure \ref{fig:percentage_frequent}. We use the identifiers defined in Table \ref{table:quality_issues} to refer to the quality issues. The top three issues with the highest percentage are \issue{I1.4, I6.2, I1.2}. Two out of the three most frequent quality issues are problems with evaluability. The three least frequent quality issues are \issue{I4.2, I4.3, I3.1}. Two out of them are issues with efficiency.

\begin{tcolorbox}
\textbf{Key findings:} Evaluability is the most frequently problematic quality dimension. Practitioners often spend a lot of time debugging MLSSs when they have external data providers. Efficiency issues are less often experienced according to our practitioners' experience. 
\end{tcolorbox}

\subsection{RQ3: How are the quality issues currently handled by the practitioners?} \label{RQ3}

In this section, we share the strategies the practitioners designed to handle the quality issues mentioned in Section \ref{RQ1}. These strategies do not solve the issues, they only attenuate their consequences. Some issues are not covered in the following, since the practitioners from the interviews and questionnaire did not share any particular approach to address them. We provide a simplified description of the strategies used by the practitioners in Table \ref{table:quality_issues_sol}. After presenting the mitigation strategies per quality aspect, we highlight in Section \ref{sec:no_mitigation} the issues for which our practitioners have not mentioned any mitigation technique.

% ----------------------------------------------------------------
%                      Quality issues sol table
% ----------------------------------------------------------------
% \setlength{\tabcolsep}{8pt} % horizontal spacing, Default value: 6pt
\renewcommand{\arraystretch}{1.5} % vertical spacing, Default value: 1
 \begin{table}
 \caption{List of techniques used by our practitioners} to mitigate the quality issues' consequences
 \centering
 \begin{tabular}{|>{\centering}m{1.5em}|>{\centering}m{2.5em}|m{30em}|}
 \hline
 & Issue's Id & \multicolumn{1}{c|}{Practitioners' mitigation techniques}   \\
 \hline
 \multirow{1}{*}[-0.5em]{\rotatebox[origin=c]{90}{\parbox[c]{1cm}{\centering Evalua-bility}}}
    & \multirow{2}{*}[-0.7em]{1.1}  
        & Evaluate the model online using a business metric.  \\
        && Evaluate the model using the feedback of domain experts, users of the MLSS, or business stakeholders.  \\
    \hline
  \multirow{1}{*}[.3em]{\rotatebox[origin=c]{90}{\parbox[c]{1cm}{\centering Explaina-bility}}}
    & 2.1 2.2 &  Instead of explaining the predictions of a model, make the system more transparent to the user. Custom strategies can be developed for that purpose (e.g. presenting historical performance on similar cases to the user).\\ 
    \hline
 \multirow{2}{*}[-1.8em]{\rotatebox[origin=c]{90}{\parbox[c]{1cm}{\centering Debugg-ability}}}
    &   &  \\[.005em]
    & \multirow{1}{*}[-0.5em]{3.1}  &  Temporarily save a copy of a record in a database whenever it has been used by a model (for training or inference). \\ 
    &   &  \\[.005em] 
    
    \hline
 \multirow{3}{*}[-4.5em]{\rotatebox[origin=c]{90}{Efficiency}}
    & 4.1 & Train models across different machines at the same time using distributed frameworks. \\
    \cline{2-3}
    & 4.2 & Use ML architectures with a reduced number of parameters. \\
    \cline{2-3}
    & \multirow{1}{*}[-4em]{4.3}  
        & Distribute computations across many computers using distributed frameworks.  \\
        && Use ML architectures with a reduced number of parameters. \\
        && Optimize data processing pipelines (e.g. by using vectorized operations). \\
        && Decommission features that are not important for prediction to reduce the load on the system and make it more efficient. \\
     \hline
 \multirow{3}{*}[-1.8em]{\rotatebox[origin=c]{90}{Maintainability}}   
    & \multirow{2}{*}[-.5em]{5.1}  
        & Follow rigorous project management methodology where data documentation and understanding play a central role. \\
        && Use a data storage procedure that forces its users to document data when entered into the system. \\ 
    \cline{2-3}
    & 5.2  & Use experiment tracking tools, namely MLflow, to automatically document the hyperparameters used to train ML models. \\
    \cline{2-3}
    & 5.3  & Share the same development environment between developers and use that environment in production. \\ 
    \hline
 \multirow{3}{*}[-6em]{\rotatebox[origin=c]{90}{Reliability}}   
    & \multirow{2}{*}[-0em]{6.1}  
        &  Automatically retrain ML models on the latest data. \\
        &&  Define a systematic data collection process to avoid annotation drift. \\
    \cline{2-3}
    & \multirow{5}{*}[-2em]{6.2}  
        &  Before using a data source, assess its reliability and, if it is not reliable, avoid using it. \\ 
        &&  Fetch the same information from different data providers to have data redundancy. \\
        &&  Monitor the data sources to address any issue in the MLSS as fast as possible. \\
        &&  In case of an issue with a data source used at inference, inform the user that the quality of the model's predictions will be degraded. \\
        &&  In case of an issue with data used for training, roll back to a previous version of the model. \\
    \hline                    
\end{tabular}
\label{table:quality_issues_sol}
 \end{table}

% ------------------------------------------------------------------
%                           Evaluability
% ------------------------------------------------------------------
\subsubsection{Evaluability} \label{RQ3:evaluability}

As explained in Section \ref{RQ1}, evaluating the quality of models offline on a test dataset with an ML performance metric (e.g., accuracy) often gives poor measures of quality \issue{I1.1}. To mitigate this problem, practitioners try to evaluate their models as frequently as possible with better proxies of quality. If the model can be evaluated online (i.e., deployed in a production environment), then the model is evaluated in a production environment using a business metric \practitioner{P1, P7, P9}. For example, for a trading system, one could measure how profitable the new system is compared to the previous one. If deploying the model in a production environment is not possible, then the model can be evaluated offline using the judgment of domain experts, users of the MLSS, or business stakeholders \practitioner{P3, P13, P23}. We refer to anyone with significant experience on a problem as domain experts. On the topic of model evaluation with domain experts, \practitioner{P3} says: \citepr{It's really important to have [...] someone who can challenge [your] results and who [...] knows the business. It's important to involve them as early as possible}. 

As mentioned in Section \ref{RQ1:evaluability}, the goal that is targeted when an ML metric such as accuracy is used is different from what is important for the end application, partially causing issue \issue{I1.1}. Hence, \practitioner{P9} suggested integrating the business perspective into the loss function of a model, so the model is optimized for what is important for the end application. For example, for a trading system, the model could be directly optimized for profit instead of accuracy. However, the practitioner only suggested that as an idea and did not implement it.

% ------------------------------------------------------------------
%                           Explainability
% ------------------------------------------------------------------
\subsubsection{Explainability} \label{RQ3:explainability}

As mentioned in Section \ref{RQ1:explainability}, one of the goals of explainability is to increase the users' trust in an MLSS's predictions by providing explanations of a model's predictions \citep{DBLP:journals/corr/abs-1909-06342}. However, we found two issues hindering transparency, namely issues \issue{I2.1, I2.2}. To address this problem, \practitioner{P23, P32} developed custom strategies to justify the model's predictions. For example, for a house price prediction system, \practitioner{P32} showed the user the price of similar houses for sale on the market. As a result, the user was able to compare the prediction of the model with similar data and decide whether the prediction could be trusted or not. More generally, \practitioner{P23} recommends presenting historical performance on similar cases to the user. For example, one could show the user how many times the model was accurate on a similar example. As a concluding thought on replacing explainability tools with custom strategies, \practitioner{P23} said: \citepr{[The custom strategies] will probably solve the transparency problem [...] even though the model won't be [able to] explain. So you're really making the system usable more than you are solving the models' explainability problem}.

% ------------------------------------------------------------------
%                           Debuggability
% ------------------------------------------------------------------
\subsubsection{Debuggability} \label{RQ3:debuggability}
In the following, we present possible remedies practitioners shared with us to address the issue \issue{3.1}. As expressed in Section \ref{RQ1:debuggability}, some practitioners have issues debugging their MLSSs, because of unstable data sources that return different values for the same queried record. This makes reproducing a bug very hard and sometimes impossible. To address that problem, \practitioner{P16} temporarily saves a copy of a record in a database whenever it has been used by a model (for training or inference). Similarly, \practitioner{P25} mentioned trying to use data version control systems to keep a history of the data used to train her models. However, the person abandoned the tool after a few weeks, because using the tool required too much effort, and the management team did not see any value in using it.

% ------------------------------------------------------------------
%                           Efficiency
% ------------------------------------------------------------------
\subsubsection{Efficiency} \label{RQ3:efficiency}
In Section \ref{RQ1:efficiency}, we explained the quality issues \issue{I4.1, I4.2, I4.3}. To reduce the time taken to train a model and partially address issue \issue{I4.1}, \practitioner{P8} used Spark's ML library to train models across different machines at the same time. When facing memory-constrained environments, \practitioner{P35} used ML architectures with a reduced number of parameters, such as MobileNet \citep{howard2017mobilenets}, to address issue \issue{I4.3}. Practitioners have reported different strategies to address latency issues with MLSSs \issue{I4.2}. \practitioner{P16, P36}'s teams typically try using a model with a reduced number of parameters, so the inference is faster. For example, \practitioner{P16}'s team focuses on tree-based models instead of deep neural networks. To avoid latency issues caused by heavy traffic, \practitioner{P36}'s team uses frameworks such as PySpark to distribute computing across different machines. \practitioner{P36, P38} mentioned addressing latency issues by optimizing the data processing pipelines. For example, loops can be replaced by vectorized operations to make the code faster \practitioner{P38}. Finally, \practitioner{P16}'s team decommissions features that are not important for prediction to reduce the load on the system and make it more efficient.

% ------------------------------------------------------------------
%                           Maintainability
% ------------------------------------------------------------------
\subsubsection{Maintainability} \label{RQ3:maintainability}
In Section \ref{RQ1:maintainability}, we explained the quality issues \issue{I5.1, I5.2, I5.3}. As expressed in issue \issue{I5.1}, maintenance is challenging when no proper dataset documentation is available. As a consequence of this problem, \practitioner{P1, P2, P19, P29} had to communicate with domain experts who can help them understand the dataset at hand, which can be time-consuming, as explained in Section \ref{RQ1:maintainability}. To avoid that issue altogether, practitioners shared strategies used by their team to keep data well documented. \practitioner{P38}'s team followed a methodology for project management in Data Science, namely TDSP\footnote{\url{https://learn.microsoft.com/en-us/azure/architecture/data-science-process/overview}}. Data understanding and documentation play a central role in the methodology, ensuring their data is sufficiently documented when developing (and maintaining) MLSSs. \practitioner{P8} built a data storage procedure that forced her team members, when entering a new dataset or data source in a database, to document data to the best of their abilities. Practitioners were asked, for example, to provide the estimated level of data quality of a data source. However, the practitioner's team ended up giving up on that solution because it was too much of a burden. 

As expressed in issue \issue{I5.2}, model maintenance is challenging when no proper model documentation is available. \practitioner{P25} uses experiment tracking tools, namely MLflow\footnote{\url{https://mlflow.org/}}, to document the hyperparameters used to train models. \practitioner{P20} considers using ML platforms with embedded model registries, namely Vertex AI\footnote{\url{https://cloud.google.com/vertex-ai}}, to be aware of the models already available inside her company. As a result, her team would avoid re-implementing models that already exist for a problem.

As expressed in issue \issue{I5.3}, managing the dependencies of an MLSS is challenging and error-prone. To avoid issues with dependencies, \practitioner{P10}'s team shares the same development environment. As a result, less effort is spent on having a functional environment. This environment is used in production as well, ensuring the developed solution is functional upon deployment. Similarly, \practitioner{P38}'s team avoids configuring and managing environments by using the ones provided and managed by their cloud provider.

% ------------------------------------------------------------------
%                           Reliability
% ------------------------------------------------------------------
\subsubsection{Reliability} \label{RQ3:reliability}
In Section \ref{RQ1:efficiency}, we explained the quality issues \issue{I6.1, I6.2, I6.3}. To address concept and data drift (issue \issue{I6.1}), our interviewees mentioned monitoring their models or data sources (e.g. \practitioner{P1, P13, P15, P20, P21, P24, P36}). However, monitoring and model retraining is not always sufficient to ensure the quality of the model's predictions. Hence, when possible, practitioners shared solutions to minimize data and concept drift. As we explain in Section \ref{RQ5}, data and concept drift can be caused by a change in the data collection process. When data collection is manual, data and concept drift can be attenuated by standardizing the data collection process and providing clear guidelines to the people collecting data \practitioner{P16, P37}.

Practitioners shared different strategies to handle unstable external dependencies (issue \issue{I6.2}). A rather straightforward one is to simply avoid using an external data source if it is an unreliable one \practitioner{P4, P10, P11, P15}. Before adding a data source to an MLSS, \practitioner{P11} assessed the reliability of the data source. As she mentioned in the following: \citepr{So when we start a project, we [...] include new datasets [only] when we are really sure that they're gonna come from a steady source. [...] You have to make sure that you're able to feed this model [...] in a consistent manner. So we just try to be very careful with the data we include in the model. This is the main solution we have}. When practitioners could not avoid using a data source, they adopted different strategies to mitigate the risks with reliability. \practitioner{P7, P10} fetched the same information from different data sources to have data redundancy. Hence, they are able to verify the correctness of the values returned by the data sources (by comparing them) and to ensure data availability at inference time. \practitioner{P7, P8, P15, P20} mentioned monitoring their data sources, doing data validation, and sending alerts in case of issues with the data sources. As a result, engineers are able to quickly detect and address a problem with data sources. However, this does not guarantee the end user will not be impacted by unreliable data sources. Hence, when a problem with a data source happens and is detected, practitioners have different strategies to attenuate the consequences. In one of her projects, \practitioner{P10} informed the user whenever a model's prediction risked being worse than usual because of an issue with a data source. Finally, when the latest training data of a model had poor quality, \practitioner{P21}'s team rolled back to the previous version of a model until the issue with the data source was resolved.

\subsubsection{Quality issues without mitigation strategies}\label{sec:no_mitigation}

\begin{table}
\centering
\begin{tabular}{|l|c|}
\hline
\multicolumn{1}{|c|}{Quality aspect} & 
\multicolumn{1}{c|}{Issues' Id}    \\
\hline
\hline
Evaluability & I1.2, I1.3, I1.4   \\
\hline
Debuggability & I3.2, I3.3   \\
\hline
Reliability & I6.3   \\
\hline
\end{tabular}
\caption{List of quality issues without mitigation techniques}\label{no_mitigation_strategy}
\end{table}

While most practitioners shared the strategies they followed to mitigate the consequences of the quality issues they experienced, a few issues were left without solutions. For clarity, we display them in Table \ref{no_mitigation_strategy}. The quality aspect with the highest number of issues without mitigation technique is evaluability. Indeed, 3 out of 4 issues are left without mitigation techniques. For practitioners, this can be problematic, since evaluability is also the quality aspect that is the most common according to their experience  (Section \ref{RQ2}). It is interesting to notice that 4 out of the top 5 most common issues (according to our results in Section \ref{RQ2}) do not have any mitigation strategy \issue{I1.4, I1.2, I1.3, I3.3}. It highlights that these issues are challenging for the practitioners we interviewed as well as the ones who answered our questionnaire. After consulting the literature, we suggest mitigation strategies for these issues along with future work directions in Section \ref{recommendations}.

\subsection{RQ4: Which data quality aspect is the most problematic in practice?} \label{RQ6}
\begin{figure}[t]
 \centering
 \includegraphics[width=0.85\textwidth]{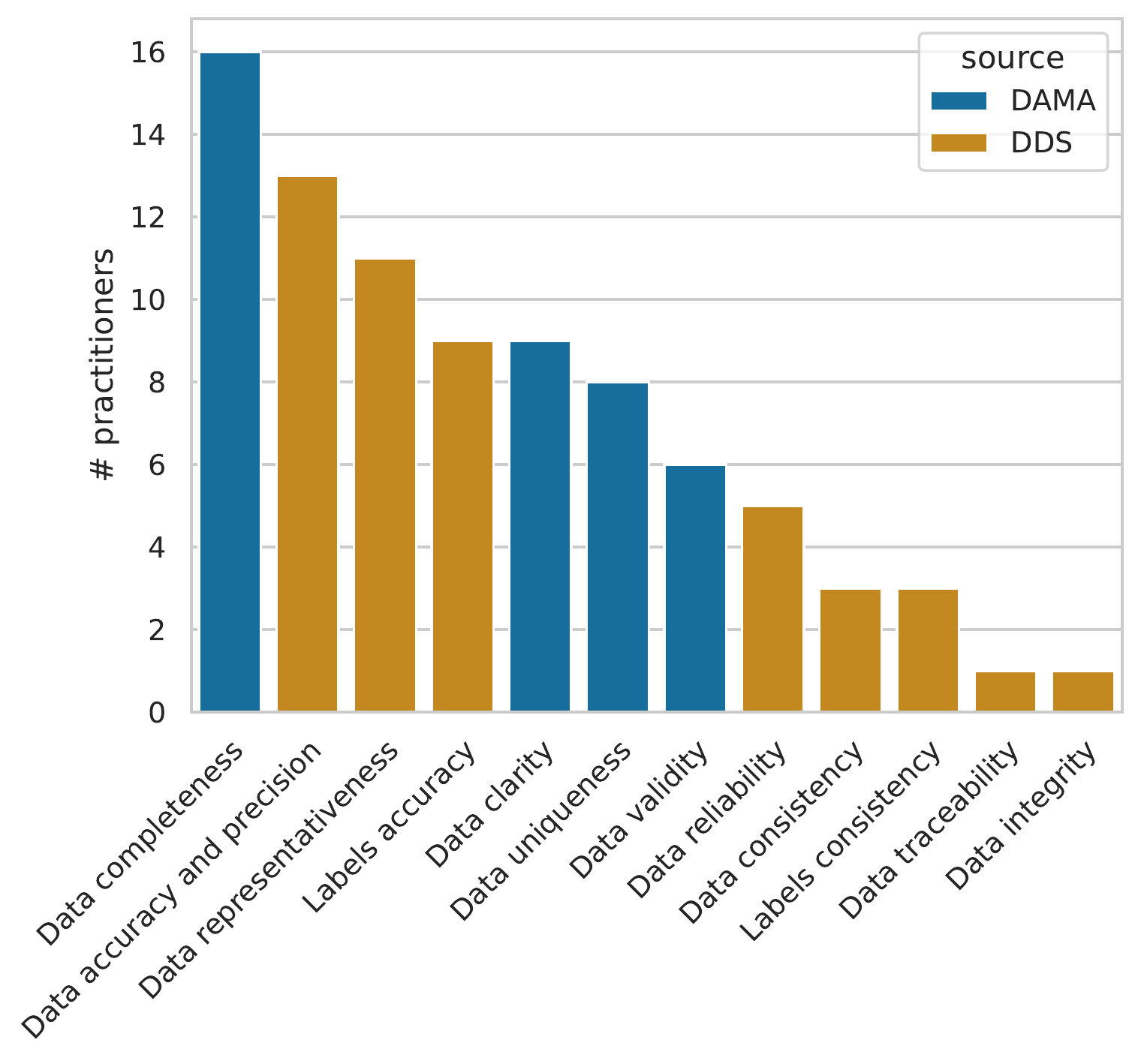}
 \caption{Number of practitioners having an issue affecting one of the quality dimensions defined by DDS \citep{cappi2021dataset} or DAMA \citep{dama}}
 \label{fig:quality_dimensions}
\end{figure}

As mentioned in Section \ref{analysis-plan}, we used the Dataset Definition Standard (DDS) \citep{cappi2021dataset} to group the data issues. A data issue is included in a data quality aspect as defined in DDS if it hinders that aspect. For example, mentions of mislabeled data are included in the \enquote{Labels accuracy} aspect. The issues that could not be included in any of DDS's data quality aspects were categorized by a researcher into one of the Data Management community's (DAMA) \citep{dama} twelve most common data quality dimensions. We detail in the annex (Section \ref{annex:RQ6}) the data issues included in each data quality aspect experienced by practitioners. We also define each of the data quality aspects or dimensions in the annex.

Figure \ref{fig:quality_dimensions} displays the number of practitioners having faced an issue with one of the quality aspects defined by DDS \citep{cappi2021dataset} or DAMA \citep{dama}. As we can observe from the figure, having accurate and complete data is challenging for practitioners. Indeed, \enquote{Data completeness}, \enquote{Data accuracy and precision}, and \enquote{Labels accuracy} are respectively the first, second, and fourth quality aspects with the highest count of issues. Data representativeness comes third in the figure, because of the large number of practitioners struggling with data and concept drift. The least affected quality aspects are \enquote{Data integrity} and \enquote{Data traceability} with only 1 practitioner having struggled with these aspects. Three of DDS's data quality aspects are not displayed in this figure, because we have not found any issue with them in our interviews. These are: \enquote{Unintended bias in data}, \enquote{Absence of Annotation bias}, and \enquote{Independence between datasets}. The data quality dimensions from DAMA we used to complement DDS's data quality aspects are \enquote{Data completeness}, \enquote{Data clarity}, \enquote{Data uniqueness}, and \enquote{Data validity}, and we counted respectively 16, 9, 8, and 6 practitioners having an issue with the respective data quality dimension.

\begin{tcolorbox}
\textbf{Key findings:} Practitioners are often faced with datasets with missing values. Additionally, they struggle to have precise and accurate features or labels. Having a dataset that is not representative of the modeled process is a recurrent issue.
\end{tcolorbox}

\subsection{RQ5: In the case of data, which data types and collection processes are the most challenging in terms of data quality?} \label{RQ4}
We answer this RQ in two parts. In Section \ref{RQ4:data-types}, we show, for each data type, the data quality issues our practitioners experienced. We do the same for data collection processes in Section \ref{RQ4:data-collection}.

\subsubsection{Data types} \label{RQ4:data-types}
Nine practitioners mentioned having issues with tabular data and six with image data. There was no issue we could confidently assume to belong to other data types. For tabular data, the problematic data quality aspects were: data accuracy and precision \practitioner{P7, P25, P32}, labels accuracy \practitioner{P32}, data integrity \practitioner{P29},  data inconsistency \practitioner{P21}, and data reliability \practitioner{P13, P15, P34, P35}. For image data, the problematic data quality aspects were: label consistency \practitioner{P21, P35} and label accuracy \practitioner{P25, P32, P40}.

\subsubsection{Data collection processes} \label{RQ4:data-collection}
In our interview, we observed shared issues between the data collection processes depending on whether they were automated or not. Hence, we present manual data collection and automated data collection separately. With manual data collection ensued data quality issues in terms of data accuracy and precision \practitioner{P7, P31, P38}, label accuracy \practitioner{P14, P40}, and label consistency \practitioner{P21, P37, P43}. For automated data collection processes, we only found issues with data accuracy and precision \practitioner{P7, P8, P20}. Additionally, automated data collection processes occasionally faced reliability issues \practitioner{P4, P7, P10, P15, P21, P40, P43}. In the following, we detail the issues with manual and automated data collection.

\paragraph{Manual data collection}
The practitioners have mentioned that manual data collection can take a lot of time \practitioner{P7, P8, P32, P38, P43}. As a result, building a dataset for ML and making sure it stays relevant (when there is data and concept drift) can be difficult. We have observed three issues with manual data collection that leads to poor data quality. First, humans occasionally make mistakes. For example, a person collecting data could misspell a word \practitioner{P7, P31}, inadvertently paste a wrong value \practitioner{P38}, or misinterpret an image and label it incorrectly \practitioner{P14, P40}. As a result, practitioners may work with data that make no sense \practitioner{P10} and spend a lot of time cleaning the dataset. Second, humans interpret events differently, which leads them to annotate data differently \practitioner{P21, P37, P43}. As illustrated by \practitioner{P37}, the same temperature can be hot for someone, but mild for someone else. Third, humans write down data differently, which can lead to some inconsistency in the data. For example, data collectors may use different conventions to represent time or use commas instead of periods to separate decimals from integer numbers \practitioner{P38}. Additionally, synonyms can also be used to identify the same object \practitioner{P25, P31}.

\paragraph{Automated data collection}
Compared to manual data collection, automated data collection has the advantage of being systematic by default. Hence, most of the issues that have been observed with manual data collection were not observed in automated data collection processes. As a result, data collection processes tend to be automatized and run periodically when possible. However, these processes may have reliability issues. Notably, practitioners have mentioned issues with web-scrapping for which the slightest change to the web page might break the process \practitioner{P4, P10, P15, P43}. As pointed out by \practitioner{P15}, maintaining web-scrapping processes is time-consuming. Sensors used for data collection may experience technical issues and be unable to send accurate data \practitioner{P7, P8, P20}. Batch processing scripts may also fail, as practitioners \practitioner{P7, P21, P40} experienced. In \practitioner{P21}'s case, it took several days to realize there was an issue since the process was not monitored. As a result, the model got trained on old data only.

\begin{tcolorbox}
\textbf{Key findings:} Manual data collection can lead to data quality issues if the process to collect data is not systematic and properly defined. On the other hand, automatic data collection processes usually have reliability issues.
\end{tcolorbox}

\subsection{RQ6: What are the challenges of data quality assurance during model evolution?} \label{RQ5}

As it is widely known in the ML community, ML models are highly sensitive to the quality of data. As a result, during the maintenance of an MLSS, a significant amount of effort must be spent on data sources in order to ensure the system's reliability. In our study, we observed three data issues practitioners have to take into consideration during the maintenance of MLSSs. First, the problem that is modeled may change over time, which will render the deployed model useless. Second, the data collection process may change over time, which, similarly, causes a drift in the distribution of data and may render the model useless. We refer to that problem as annotation drift. Third, external data providers may not have the standards of reliability that are required to build MLSSs. As a result, practitioners must put extra care when dealing with external providers in order to ensure the reliability of their MLSSs.

\subsubsection{Data and concept drift} \label{RQ5:concept-data-drift}
Data drift refers to a change in the distribution of records over time. When modeling the function F that maps X (i.e., records) to Y (i.e., labels), data drift is a change in P(X) over time. Concept drift refers to a change in the function F over time \citep{tannor_2023}. Concept and data drift are problems that must be considered by practitioners, or else the deployed model will lose predictive capabilities and become stale. According to our interviews, practitioners are already well aware of these problems and automatically retrain their models to address the issue. However, we observed two scenarios where retraining was not sufficient to attain past performances. First, sudden changes in the data, such as pandemics like COVID-19, render the past data useless for the new prediction problem. As a result, obtaining the same level of performance as with the previous models may be almost impossible, since not enough data is collected for the new problem \practitioner{P9, P20}. Speaking of her experience with COVID-19, \practitioner{P20} said: \citepr{We had this weird time where we know the data has changed, but we don't have enough data to train new models because it hasn't been that long. We kind of just have to [wait] until enough time goes by and we can start retraining these models}. Second, performance issues caused by concept drift can not always be solved by retraining the model \practitioner{P15}. Concept drift can render features useless. If a significant number of features are affected, then predicting the target variable will be a much more difficult task for the ML model. Hence, periodically retraining a new model and deploying it will not be sufficient to preserve ML performance.

\begin{tcolorbox}
\textbf{Key findings:} While data and concept drift are well-known by practitioners, they are still considered hard problems. Automatically retraining models is not a panacea to these issues. Performance drops because of data or concept drift are sometimes inevitable due to insufficient data for retraining.
\end{tcolorbox}

\subsubsection{Annotation drift}
While it is clear that data and concept drift may be caused by a change in the modeled problem (e.g., what happened for many MLSSs when COVID-19 began), it is maybe less obvious that data collection may generate the same problem. For simplicity, we refer to drift in the annotation process as annotation drift. As mentioned in Section \ref{RQ4}, manual data collection is not always a systematic process, and, as a result, is error-prone. Humans interpret events differently and may write down information differently. Hence, annotation drift can be observed if the process for collecting data is not described exhaustively. Annotation drift naturally happens as people change, or as team members are replaced \practitioner{P36}. Drastic changes in the data distribution may happen, if, for example, the annotation procedure is changed \practitioner{P24, P25, P36}. In this case, one should define a procedure for transforming the data collected with the old standard into the new standard. If it is not done, a lot of data will be lost, which might hinder the training of ML models, as experienced by \practitioner{P25}.

\begin{tcolorbox}
\textbf{Key findings:} Data and concept drift sometimes creep from unexpected places, such as the data collection process. Annotation drift happens when no systematic data collection process is defined.
\end{tcolorbox}

\subsubsection{Unreliable external data providers}
% Definition of the problem
Practitioners revealed that external data sources are often the cause of many issues during maintenance. We consider any data source to be external if it is not managed by the team building the model. Because data providers are not aware that their consumers are ML models, maintainers of MLSSs face a lot of reliability issues. On the topic of the reliability of third-party APIs, \practitioner{P23} said: \citepr{[...] it's a nightmare. [..] their job is not to provide ML-ready data. So the data is not formatted, [and require] quite significant post-processing}. She also added: \citepr{They often sell their data to people who don't require high uptime [...] which means that sometimes it'll break. And most of their clients don't care. But if you're trying to build an AI system, you do care quite a lot}. Overall, we observed two problems with external data providers. First, practitioners may have unreliable access to data. Second, data providers may return inconsistent information. That is, the same information accessed at different times might have different values.

% Issue 1 v2
We observed two causes of accessibility issues with data providers. First, they may be unable to deliver data at times because of technical issues \practitioner{P10, P11, P21, P23}. For example, their API may be down. Second, they may unexpectedly change the definition of their interface \practitioner{P15, P27}. As a result, the affected MLSSs are not able to fetch data until the code used to communicate with the API is adapted. 
Not being able to access data is a problem at any time in the model life cycle. At inference time, the model will not be able to make a prediction unless replacements for the missing values are found. The replacements are not guaranteed to be accurate; thus hindering the quality of predictions of the model. At training time, important information may be lost if records with missing values are thrown out of a dataset.

% Issue 2
Aside from occasional accessibility issues, data sources may serve inconsistent information. More precisely, the same information fetched at different times may have different values. This is a problem for an ML model since it means it has learned from data that is not representative of the new data distribution. A data provider may return different values for the same records for various reasons. In our interviews, we have seen two. First, the data providers may choose to change the units they have used to measure some information \practitioner{P4, P6, P43}. For example, dates in British format could suddenly be changed to Unix data format \practitioner{P43}. Second, the data providers may voluntarily alter historical data \practitioner{P4, P6, P10, P12}. This may happen if the provider believes that the data they previously served was inaccurate. For example, practitioner \practitioner{P4} consumed data from a weather API. One day and without notice, the data providers corrected old weather recordings that were faulty because of a broken sensor.

\begin{tcolorbox}
\textbf{Key findings:} Ensuring the reliability of an MLSS when data is collected from external data providers is challenging for practitioners. External data providers may be unavailable at times, or serve inconsistent information.
\end{tcolorbox}

\section{Recommendations} \label{recommendations}
In this section, we provide a set of recommendations for practitioners to help them tackle the quality issues mentioned in Section \ref{RQ1} and the data quality issues covered in Section \ref{RQ6}. We do not repeat the ones that have already been mentioned in Section \ref{RQ3}. Additionally, based on our results, we provide some directions for future work. Our first suggestion of future research directions is to measure how common the mitigation techniques described in Section \ref{RQ3} are and how effectively they can address quality issues. A similar research work could be done with the mitigation techniques suggested in this section.

\subsection{Data quality} 
As observed in Section \ref{RQ6}, the data quality issues that were the most mentioned in our interviews were incorrect feature values and labels. Fixing these kinds of errors often consumes a lot of time \practitioner{P8, P3} and is often done manually \practitioner{P9}. According to our interviews, there is not a wide variety of tools that are used by practitioners to clean data. \practitioner{P6} used Data Wrangler \citep{hudgeon_nichol_2020} to fasten the data cleaning process. However, that tool is limited to one cloud provider; thus limiting its usage. Hence, as future work, we recommend developing tools to facilitate the process of detecting and repairing errors in data. Similarly, we recommend developing advanced tools for data cleaning of duplicates and missing values, since a large number of practitioners had issues with it (see Section \ref{RQ6} for further details). 

As mentioned in Section \ref{RQ6}, we used DDS \citep{cappi2021dataset} to define data quality aspects. Because the DDS is not exhaustive, some quality issues could not be categorized in any quality aspect provided in DDS. Thus, we completed DDS's definition of data quality with the twelve most common quality dimensions defined by DAMA \citep{dama}. To answer our RQs, using the later work would not have been preferable either. Some quality aspects defined in the DDS were not covered by DAMA, because the context of the latter was not ML. For example, ``absence of annotation bias" is considered in DDS but not in DAMA. Hence, because either work could not capture every data issue we observed, we recommend future works to develop a taxonomy of data quality for ML applications.

\subsection{Evaluability}
As discussed in Section \ref{RQ1:evaluability}, one of the reasons why offline evaluation of models is inaccurate is because the goal targeted by accuracy is different from what is important for the user. We provide in Section \ref{RQ3:evaluability} a set of alternatives to evaluate models to have more accurate measures of quality. As observed by \practitioner{P7} in a follow-up conversation after the interview, junior practitioners rely too much on accuracy to evaluate their model and end up deploying models that perform poorly in production. We believe that a root cause for that problem is the ubiquity of these metrics in the curriculum of students. We encourage educational programs to include techniques to evaluate a model in the context of building MLSSs. Future works could look into building loss functions that include the business perspective, as \practitioner{P9} suggested. For example, for a financial trading system, the model could be rewarded based on how profitable a decision was. Training a model using these loss functions would create models that are already optimized for the end application. Alternatively, future works could look into techniques to post-process ML models to optimize the business need. 

As explained in Section \ref{RQ1:evaluability}, defining a good business metric to evaluate an MLSS can be difficult for some applications, such as recommender systems (issue \issue{1.2}). In these scenarios, practitioners may discuss with business stakeholders to conjointly develop business metrics that accurately measure how useful the model is for the application.

In Section \ref{RQ1:evaluability}, we described the problem of dataset evaluability (issue \issue{I1.4}). Evaluating the quality of a dataset (or knowing its potential issues) is difficult. Hence, we propose two approaches for practitioners to get an estimation of data quality faster. First, data visualization tools, such as Facets\footnote{\url{https://pair-code.github.io/
facets/}}, can be used to get a better understanding of the dataset at hand and maybe understand quality issues easier. The data visualization tools can be used conjointly with tools to automatically generate interesting visualizations such as SeeDB \citep{vartak2015seedb}. Second, error detection tools on datasets can be used to get an approximate count of the number of errors in a dataset. We recommend practitioners to explore unsupervised error detection tools, since they do not require a significant time investment before they can be applied on a dataset. Notably, \citet{narayan2022can} showed that pre-trained transformer models could detect errors in tabular data without any training examples. The models achieve even better performances when given training examples. With only 10 labeled samples, their approach surpassed state-of-the-art error detection approaches. \citet{liu2022picket} analyzed the reconstruction error of an auto-encoder model to detect dirty samples. Samples that the auto-encoder struggles to reconstruct are marked as being dirty. 

One of the reasons to evaluate a dataset's quality is to estimate the feasibility of developing an ML solution using the dataset at hand. When asked about her biggest pain points, \practitioner{P3} answered the following: \citepr{the uncertainty of [knowing whether the ML solution will] work or not;  maybe speeding up the process? [...] getting to a point where [I] can assess the quality of an approach faster, that would help}. We propose two approaches for practitioners to get that information without measuring data quality. First, AutoML techniques \citep{doke2021survey} can be used to get the ML performance of a model on the dataset with minimal effort. Second, Bayes error rate estimators, such as the work of \citet{renggli2022snoopy_system}, can be used to directly estimate the feasibility of reaching a certain performance level on the dataset at hand.

\subsection{Explainability}
Issue \issue{I2.2} refers to the unreliability of explanation techniques. As expressed by \citet{https://doi.org/10.48550/arxiv.1811.10154}, explainability techniques only are approximations of the underlying decision process of a model. Hence, they are bound to be wrong and they can not be relied on. \citet{merrick2020explanation} shows different situations where Shapley value based feature attribution methods produce faulty explanations. Similarly, \citet{laberge2022fooling} shows how these feature attribution methods can be manipulated with adversarial attacks to reduce the importance of targeted features. Common feature attribution techniques, such as the work done by \citet{ribeiro2016should}, use correlation to explain a model's prediction. They do not capture the causal influence of an input on a model's prediction. Therefore, an interesting research direction is to develop causal explanation techniques. As pointed out by \citet{DBLP:journals/corr/abs-1909-06342}, there are very few numbers of studies conducted in that direction \citep{pmlr-v97-chattopadhyay19a}.

\subsection{Debuggability} \label{discussion:debuggability}
Issue \issue{I3.1} refers to the challenge of debugging MLSS with unstable data sources. Most specifically, debugging an MLSS is difficult when two queries to an external data source for the same information return different values (see Section \ref{RQ1:debuggability} for more details). In the following, we refer to this problem as inconsistent information. In Section \ref{RQ3:debuggability}, we shared the solution of a practitioner to address this problem. She temporarily saved any data ingested by her models, so it could be fetched later on for debugging. However, this solution has its shortcomings. First, this approach requires saving a large volume of data. Second, the older bugs can not be reproduced, since the data is eventually deleted. Hence, we highlight the usefulness of recommendations given for reliability issues in Section \ref{RQ3:reliability} to address the debuggability issue \issue{I3.1}. By applying these recommendations, the data sources become more reliable, hence avoiding the issue \issue{I3.1}. 

Issue \issue{I3.2} refers to the challenging aspect of debugging data streaming systems: picturing what data should look like after a series of operations is difficult. As we explained in Section \ref{RQ1:debuggability}, one might compare data under test against the mean or the mode to detect abnormalities. However, according to our interviewees, this is not easily doable, as \practitioner{P17} said the following: \citepr{So if you want to inspect the middle steps [of a data streaming pipeline], you have to [do the] effort [of] sending all the data to another table. [...] After this [job] is done, then [you can have] a look at the table [and compare your values]}. To avoid this long process when debugging data streaming pipelines, \citet{rezig2020dagger} proposes Dagger, a data debugger that can automatically provide the value of any tracked variable across any step of a data pipeline. Compared to traditional code debuggers, the tool proposes features to make data debugging easier, such as a ``compare" command that enables comparing the value of a variable at a specific time across different runs of the pipeline (i.e., for different data records). To efficiently save the value of a record throughout a data pipeline, only the changes to a value are saved. We recommend future works to develop data debuggers that can: (1) efficiently save the value of a record throughout a pipeline, (2) provide features to easily compare data values across data pipeline runs (e.g., getting the mean), and (3) easily append the debugger to existing data streaming system.

Issue \issue{I3.3} refers to the challenging aspect of debugging MLSSs when its data sources are managed by external data providers. As explained in Section \ref{RQ1:debuggability}, this can easily become cumbersome because of long back-and-forth communications with the maintainers of the data source. Furthermore, the maintainers might not prioritize the issue because they have their own agenda. In the following, we provide a set of recommendations to practitioners to address that issue.
% Internal data sources
When a data source is internal to a company, the company can raise awareness among data producers on the important role they play in ensuring the quality of ML models. For example, a company could organize showcase meetings, where ML teams present their work to data producers. As a result, data source maintainers would be more inclined to help ML model maintainers when they require assistance. A similar recommendation was made in \citet{mailachsocio}. Another way to tackle the problem is to establish clear bug prioritization measures that prioritize bugs impacting the ML model, so that the data source maintainers would be incentivized to help fix the issue. Finally, when possible, similar to full-stack developers for web applications, we recommend companies to create full-stack developers roles for ML pipelines. These practitioners could have technical knowledge of the model as well as its data sources. As a result, they would be able to fix issues with data sources without relying on the assistance of the maintainers of a data source. 
% External data sources
The above recommendations can not be applied when the data source is not managed by the company running the MLSS. Hence, before adding a data source to an MLSS, we recommend practitioners to verify the quality of the customer support that the external data providers will provide in case of an issue. This assessment can be done conjointly with the one discussed in Section \ref{RQ3:debuggability} (i.e., assessing the reliability of a data source before using it).

\subsection{Efficiency} \label{discussion:efficiency}
In addition to the recommendations given in Section \ref{RQ3:efficiency}, we suggest using quantization methods \citep{gholami2021survey} to the practitioners that use neural networks in order to reduce the models' size in memory and effectively address issue \issue{I4.3}.

\subsection{Reliability}
As discussed in Section \ref{RQ1:reliability}, ensuring the reliability of an MLSS is difficult when its data sources are unreliable. We already provided a set of recommendations to address the problem in Section \ref{RQ3:reliability}. Similar to the recommendation given in \ref{discussion:debuggability}, we recommend companies to raise awareness among data producers on the important role they play in ensuring the quality of ML models. Good practices such as API versioning should be encouraged.

\section{Threats to validity}
Using some of the validity threats described in \citet{feldt2010validity}, we divide the analysis of the limitations of our study into five categories: internal validity, external validity, content validity, confirmability validity, construct validity, and conclusion limitations.

% Internal validity refers to the degree of confidence that the findings are trustworthy. 
We identified two threats to internal validity. First, as mentioned in Section \ref{metho:interview-process}, an interview was conducted in French, and the others, in English. It is possible that the questions asked when the interviews are in English are not exactly the same as the ones asked when the interviews are in French. While our primary interviewers have a good understanding of both languages, their choice of words might convey slightly different meanings, leading to bias. To mitigate that threat, the second interviewer was welcome to rephrase the question of the first interviewer if it was poorly translated. Second, there is a risk that our findings for \textbf{RQ1} are inaccurate because of our data collection method. Asking practitioners about the root cause of quality issues may lead to misunderstandings or even misinformation in case they do not want to admit their mistakes. Likewise, if the practitioners are too inexperienced, they might struggle to identify real problems and instead describe issues that are caused by their lack of experience. To mitigate this threat, we cross our findings with MoovAI knowledge on similar cases. Similar studies have included practitioners with less than three years of experience (as we did) \citep{hidellaarachchi2024impact, humbatova2020taxonomy, yang2023interview}. \\

% External validity refers to the degree the findings are generalizable \citep{feldt2010validity}. 
We identified three threats to external validity. First, as discussed in Section \ref{demographic-information}, a significant portion (40\%) of our interviewees are practitioners of MoovAI. Thus, it is possible our findings are applicable only to small companies (11 to 50 employees) whose main service is developing MLSSs for other companies (i.e., consulting). In an attempt to mitigate this bias, we interviewed a larger number of practitioners from varying company sizes. Second, there is a chance that the practitioners who participated in our study experienced quality issues in MLSSs more than the average practitioner, leading us to believe that some issues are more important than they truly are in practice. In other words, the practitioners who participated in our study could be more interested to participate than the average practitioner, because they experienced more issues than the average practitioner. To mitigate that issue, we validated our results using a questionnaire. Because the questionnaire requires less time to answer than participating in our interview, we expect that the people who answered our questionnaire are more likely to be representative of the general population of practitioners. Third, a limitation of our approach could come from keywords (tags) we use for finding candidates over Q\&A websites and social media. For keywords, we use \textit{data-cleaning, dataset, machine-learning} and \textit{artificial-intelligence} which are sufficiently general to match a lot of users and are more likely to result in False Positive (users that we would end up ignoring) rather than False Negative. \\

% Content validity refers to the degree a test is representative of all aspects of a construct \citep{Middleton_2022}.
We identified a threat to content validity. In our case, there is a risk that our interview guide does not enable us to find all the quality issues. The interview guide might not be comprehensive enough to cover every quality issue. If no practitioner mentioned an issue that was not probed in the interview guide, then we might have missed an issue. To mitigate that threat, we ask open questions so that the interviewee has many opportunities to share issues we did not suspect.

% Confirmability validity refers to the degree the findings are shaped by the participants rather than by the researchers \citep{feldt2010validity}. 
We have identified two threats to confirmability validity. First, the coding of the interview may be prone to researcher bias because it partially relies on subjective interpretation. To mitigate this bias, each interview is coded by two researchers, and inconsistencies in the codes are resolved by a third researcher during moderation. Second, there is a risk that some quality issues are over-represented in our interview guide. We are conducting semi-structured interviews; some conversations with the interviewees are improvised. Thus, it is possible that the findings are tainted by the preconceptions of the interviewer about the potential quality issues in MLSSs. For example, there could be more emphasis on explainability issues in the interview, which may lead to an over-representation of these issues in our findings. To mitigate that threat,  we used \citet{amershi2019software}'s work to ask questions covering every step of an ML workflow.

% Construct validity refers to the degree to which a test measures the construct that it is intended to measure \citep{Middleton_2022}. 
We have identified a threat to content validity. When validating our results with the questionnaire, it is possible that the practitioners misinterpret the description we provide for each quality issue. To mitigate this, the descriptions are reviewed by two researchers. Furthermore, we provide examples when the description is ambiguous.

% Could mention that depending of the state of mind of the interviewee, we get different results

% Conclusion refers to the degree to which the conclusions made by the researchers are reasonable \citep{Trochim}. 
We have identified a threat to conclusion validity. There is a risk that researchers misinterpret an issue described by a practitioner. To avoid that issue, each transcript is coded by two researchers. If an important excerpt from a transcript required further details to be understood, we contact the respective practitioner for clarification.

\section{Conclusion}
In this paper, we investigate the characteristics of real-world quality issues in MLSSs from the viewpoint of practitioners. We presented the results of the interviews we conducted with 42 practitioners and the questionnaire answered by 21 practitioners.

We extracted a list of 18 quality issues and, for every issue, we explained the causes and consequences. We measured how common the quality issues are according to the practitioners' experience. We found that the quality dimensions that are considered to be the most problematic are: evaluability (see Section \ref{RQ1:evaluability} for a definition), debuggability, and reliability. We share 21 strategies used by the practitioners we interviewed to mitigate the aforementioned quality issues. After consulting the literature, we proposed 12 other strategies. Regarding data quality, we describe the data quality issues encountered by our practitioners and how frequent they are according to their experience. Notably, we found that data completeness (missing values) and data and label accuracy are the most common data quality issues. We also explain the challenges encountered with different data collection processes that lead to data quality issues and we characterize the challenges of data quality assurance during model evolution. Finally, we propose some recommendations for future works. 

Our study shows that (1) using ML in software systems brings a unique set of challenges, and (2) practitioners struggle to build qualified MLSSs. Based on our results, we encourage researchers to develop tools to assist practitioners when building MLSSs. We hope that our findings will serve as a solid foundation for future work to come.

% Authors must disclose all relationships or interests that  could have direct or potential influence or impart bias on the work: 
\section*{Compliance with Ethical Standards}
The authors declare that they have no conflict of interest.

\section*{Data availability}
The datasets generated during and/or analyzed during the current study are available in the GitHub repository, \url{https://github.com/poclecoqq/quality_issues_in_MLSSs}

\begin{acknowledgements}
This work is partly funded by the Natural Sciences and Engineering Research Council of Canada (NSERC), PROMPT, and Les Technologies MoovAI Inc. 
\end{acknowledgements}

% BibTeX users please use one of
\bibliographystyle{spbasic}      % basic style, author-year citations
\bibliography{references}   % name your BibTeX data base

\section{Annex}
In this section, we share additional information with readers who want a more detailed description of our results. Section \ref{annex:questionnaire} and \ref{annex:RQ6} extend our answers to RQ2 and RQ4 respectively.

\subsection{Answers to our questionnaire} \label{annex:questionnaire}
We present in Table \ref{table:quality_issues_questionnaire} the raw results obtained from our questionnaire. The column N/A (i.e., not answered) refers to the number of times the question has not been answered. We identify the quality issues using the identifiers assigned in Table \ref{table:quality_issues}.

% ----------------------------------------------------------------
%                      Quality issues table
% ----------------------------------------------------------------
% \setlength{\tabcolsep}{8pt} % horizontal spacing, Default value: 6pt
\renewcommand{\arraystretch}{1.8} % vertical spacing, Default value: 1
 \begin{table}
 \centering
 \begin{tabular}{|>{\centering}m{1.3em}|>{\centering}m{1em}|>{\centering}m{2em}|>{\centering}m{1em}|>{\centering}m{1em}|>{\centering}m{1em}|>{\centering}m{1em}|>{\centering\arraybackslash}m{1em}|}
 \hline
 \multicolumn{3}{|c|}{} & \multicolumn{5}{c|}{Likert Score} \\
 \hline
 & Id & N/A & 1 & 2 & 3 & 4 & 5  \\
 \hline
 \multirow{3}{*}[0em]{\rotatebox[origin=c]{90}{ \parbox[c]{2.5cm}{\centering Evaluability}}}
        & 1.1  & 1 & 0  & 2 & 8 & 9 & 2   \\
        & 1.2  & 1 & 0  & 4 & 3 & 8 & 5   \\
        & 1.3  & 2 & 0  & 3 & 4 & 7 & 5   \\ 
        & 1.4  & 9 & 0  & 1 & 1 & 5 & 5   \\ 
 \hline
  \multirow{2}{*}[0em]{\rotatebox[origin=c]{90}{\parbox[c]{1cm}{\centering Explai-nability}}}
        & 2.1  & 0 & 1  & 3 & 6 & 7 & 4   \\
        & 2.2  & 0 & 1  & 5 & 7 & 4 & 4   \\ 
    \hline
 \multirow{3}{*}[-.5em]{\rotatebox[origin=c]{90}{ \parbox[c]{1cm}{\centering Debugga-bility}}}
        & 3.1  & 2  & 3  & 6 & 6 & 3 & 1   \\  
        & 3.2  & 3 & 2  & 5 & 7 & 3 & 1   \\
        & 3.3  & 1 & 0  & 3 & 5 & 2 & 10  \\ 
    \hline
 \multirow{3}{*}[0em]{\rotatebox[origin=c]{90}{Efficiency}}   
        & 4.1  & 0 & 1  & 5 & 8 & 3 & 4   \\
        & 4.2  & 0 & 1  & 11 & 5 & 3 & 1   \\
        & 4.3  & 1 & 3  & 8 & 5 & 3 & 1   \\
     \hline
 \multirow{3}{*}[-.50em]{\rotatebox[origin=c]{90}{\parbox[c]{1cm}{\centering Maintain-ability}}}   
        & 5.1  & 0 & 2  & 6 & 4 & 5 & 4  \\
        & 5.2  & 0 & 1  & 6 & 7 & 2 & 5  \\
        & 5.3  & 0 & 4  & 4 & 4 & 6 & 3  \\ 
    \hline
 \multirow{3}{*}[-0.2em]{\rotatebox[origin=c]{90}{ Reliability}}   
        & 6.1  & 0 & 0 & 3 & 6  & 7 & 5  \\
        & 6.2  & 0 & 3  & 1 & 3 & 12 & 2  \\
        & 6.3  & 0 & 2  & 5 & 4 & 6 & 4  \\ 
    \hline                    
\end{tabular}
\caption{Answers to our questionnaire}
\label{table:quality_issues_questionnaire}
 \end{table}

\subsection{Data quality aspects classification} \label{annex:RQ6}
In this section, we provide the details regarding how the results shown in section \ref{RQ6} were compiled. More specifically, for each of DDS' data quality aspects or DAMA's data quality dimensions, we specify the data quality issues included in the category along with the practitioners that faced these issues.

\subsubsection{DDS' data quality aspects}
\begin{itemize}
    \item \textbf{Data representativeness}: DDS \citep{cappi2021dataset} defines data representativeness as: ``[data that] contains key characteristics in proportions similar to the ones of the population". We included this category concept and data drift \practitioner{P3, P4, P8, P9, P15, P20, P25, P30, P32, P36, P43}. 
    
    \item \textbf{Data traceability}: DDS \citep{cappi2021dataset} defines data representativeness as: ``a way to maintain a connection between the different development artifacts, including requirement specifications, dataset, models, etc.". In this RQ, we are interested only in data. Hence we exclude requirements specifications and models from the data traceability definition. As mentioned in Section \ref{RQ1:maintainability}, \practitioner{P8} once could not know the provenance of data and its lineage, which, in turn, made the system difficult to maintain and unreliable. 

   \item \textbf{Data accuracy and precision}: DDS \citep{cappi2021dataset} defines data accuracy and precision as follows: ``Data are accurate if they represent with high fidelity those that will be acquired in operation. Precision refers to the closeness of data related to the same operation situation." \practitioner{P7, P8, P13, P14, P15, P20, P25, P31, P32, P35, P38, P43} mentioned facing noisy features once.
    
    \item \textbf{Data reliability}: DDS \citep{cappi2021dataset} defines data reliability as follows: ``Data are reliable if they are considered credible and relevant for the operational domain". \practitioner{P23} finds it difficult to get relevant data for ML problems. \practitioner{P38} doubt the credibility of public data sources such as Wikipedia when scraping data. \practitioner{P13, P15, P34} mentioned dealing with outliers in their datasets.
    
    \item \textbf{Data consistency}: DDS \citep{cappi2021dataset} defines data consistency as follows: ``Data consistency is [...] the absence of discrepancy between data concerning the same object (e.g., two different birthdays for the same person)." \practitioner{P7, P21} faced that issue when having to integrate data coming from different sources. \practitioner{P35} mentions that data consistency is an issue she recurrently faces.  
    
    \item \textbf{Data integrity}: DDS \citep{cappi2021dataset} defines data integrity as follows: ``Data integrity is defined as the assurance of the accuracy and integrity of data over its entire life cycle". In a past project, \practitioner{P29} could not ensure the integrity of its data because it was not version-controlled and it was shared between team members with a spreadsheet.
    
    \item \textbf{Labels consistency}: DDS \citep{cappi2021dataset} defines label consistency as follows: ``Consistency refers to the fact that the same label is assigned to the same object of interest when it is represented several times in the dataset". \practitioner{P21, P35, P37, P43} mentioned having issues with inconsistent labels.
    
    \item \textbf{Labels accuracy}: DDS \citep{cappi2021dataset} defines label accuracy as follows: ``Accuracy of labeling is defined as the correctness of the labeling with respect to the true value (or ``ground truth")". \practitioner{P14, P21, P23, P24, P25, P30, P32, P40, P43} mentioned having issues with noisy labels.
      
\end{itemize}

\subsubsection{DAMA's data quality dimensions}

\begin{itemize}
    \item \textbf{Data completeness}: DAMA defines data completeness (of data values) as follows: ``the degree to which all required data values are present". \practitioner{P2, P3, P4, P6, P8, P12, P13, P16, P19, P27, P30, P31, P32, P34, P35, P43} mentioned having a dataset with missing values.
    
    \item \textbf{Data validity}: DAMA defines data validity as follows: ``the degree to which data values comply with rules". \practitioner{P1, P3, P6, P25, P31, P38} mentioned having a dataset with values that were not formatted correctly. For example, commas could be used interchangeably with periods to separate decimals from integer numbers in a dataset.
    
    \item \textbf{Data clarity}: DAMA defines data clarity as follows: ``the ease with which data consumers can understand the metadata". As described in Section \ref{RQ1:maintainability} \practitioner{P1, P2, P5, P7, P8, P19, P24, P29, P38} used datasets with missing metadata.

    \item \textbf{Data uniqueness}: DAMA defines data uniqueness as follows: ``The degree to which records occur only once in a data file". \practitioner{P5, P19, P20, P21, P25, P31, P35, P38} mentioned handling duplicates (i.e. records that refer to the same real-world entity).
\end{itemize}

\subsection{Structured overview of quality issues} \label{annex:overview}

In this section, we provide a structured overview of the results of our study. For each quality aspect, we present a table listing the quality issues we found along with their causes, their consequences, and the mitigation strategies shared by our practitioners. While all issues harm the quality aspects they are categorized into, we do not list this as a consequence since it is implied.

\begin{table}
\centering
\begin{tabular}{|c|p{10cm}|}
\hline
Issue Id &  \multicolumn{1}{c|}{Overview of the issues}
\\ \hline
    \multirow{1}{*}[-10em]{I1.1}  & \begin{itemize}
          \item[$\bullet$] \textbf{Description}:  Evaluating the quality of a model offline (i.e., not in a production environment) is inaccurate even when the dataset used for evaluation is representative of the data distribution in production. 
          \item[$\bullet$] \textbf{Causes}: 
          \begin{itemize} 
              \item[$\circ$] The goal that is targeted when an ML metric such as accuracy is used is different from what is important for the end application.
              \item[$\circ$] The evaluation procedures do not take into consideration the system in which the model is embedded.
          \end{itemize}
          \item[$\bullet$] \textbf{Consequences}:
          \begin{itemize}
              \item[$\circ$] Time is spent on improvements that do not provide any real value.
              \item[$\circ$] Models of low quality are deployed.
              \item[$\circ$] Practitioners are unable to prove the value added by an MLSS to business stakeholders. Hence, stakeholders may lose confidence that the MLSS brings value to their company.
          \end{itemize}
          \item[$\bullet$] \textbf{Mitigation strategies}: 
          \begin{itemize}
              \item[$\circ$] Evaluate models as frequently as possible with better proxies of quality (e.g. using the feedback of a domain expert or by deploying the model in a production environment).
          \end{itemize}
        \end{itemize}     
\\ \hline
    \multirow{1}{*}[-4em]{I1.2}  & \begin{itemize}
          \item[$\bullet$] \textbf{Description}: Defining a good business metric for evaluating an MLSS is difficult. For MLSS, a business metric refers to the degree to which the MLSS successfully achieves the goal it has been built for. 
          \item[$\bullet$] \textbf{Cause}: Business metrics sometimes represent concepts that are difficult to define in mathematical terms.
          \item[$\bullet$] \textbf{Consequences}: Same as I1.1
        \end{itemize}     
        \\  
        \hline
    \multirow{1}{*}[-5em]{I1.3}  & \begin{itemize}
          \item[$\bullet$] \textbf{Description}: Trying to simulate the environment/system in which the model will operate (to evaluate the model in this simulated environment) is difficult and error-prone. 
          \item[$\bullet$] \textbf{Cause}: A model's environment is complex and may interact with the model in ways that are not obvious (e.g. feedback loops \citep{NIPS2015_86df7dcf}). 
          \item[$\bullet$] \textbf{Consequences}:
          Same as I1.1
        \end{itemize}     
        \\  
        \hline
    \multirow{1}{*}[-7em]{I1.4}  & \begin{itemize}
          \item[$\bullet$] \textbf{Description}: Evaluating the quality of a dataset (e.g., presence of incorrect labels, noisy/incorrect features, wrong format data, etc.) is difficult and time- consuming.
          \item[$\bullet$] \textbf{Cause}: Evaluating a dataset's quality requires a deep understanding of the domain and the current tools to help with that process do not provide significant help.
          \item[$\bullet$] \textbf{Consequences}:
          \begin{itemize}
              \item[$\circ$] Assessing the feasibility of applying ML to a problem is difficult.
              \item[$\circ$] Practitioners struggle to estimate the effort needed to clean a dataset.
              \item[$\circ$] Selecting a dataset without being able to properly estimate its quality is error-prone.
          \end{itemize}
        \end{itemize}     
        \\  
        \hline
\end{tabular}
\caption{Overview of evaluability issues}\label{soverview:evaluability}
\end{table}

\begin{table}
\centering
\begin{tabular}{|c|p{10cm}|}
\hline
Issue Id &  \multicolumn{1}{c|}{Overview of the issues}
\\ \hline
    \multirow{1}{*}[-7em]{I2.1}  & \begin{itemize}
          \item[$\bullet$] \textbf{Description}:  Explaining a model’s predictions to people without ML knowledge (e.g. business stakeholders, users) using explainability techniques is challenging.
          \item[$\bullet$] \textbf{Cause}: Explainability techniques require technical knowledge to be interpreted effectively.
          \item[$\bullet$] \textbf{Consequences}:
          \begin{itemize}
              \item[$\circ$] Explanation techniques are seldom used to increase users’ confidence in the MLSS.
          \end{itemize}
          \item[$\bullet$] \textbf{Mitigation strategies}: 
          \begin{itemize}
              \item[$\circ$] Instead of explaining the predictions of a model, make the system more transparent to the user. Custom strategies can be developed for that purpose (e.g. presenting historical performance on similar cases to the user).
          \end{itemize}
        \end{itemize}     
\\ \hline
    \multirow{1}{*}[-8em]{I2.2}  & \begin{itemize}
          \item[$\bullet$] \textbf{Description}:  The explanation techniques sometimes provide explanations that do not make sense and can not be relied on.
          \item[$\bullet$] \textbf{Cause}: Explainability techniques only are approximations of the underlying decision process of a model \citep{https://doi.org/10.48550/arxiv.1811.10154}. Hence, they are bound to be wrong.
          \item[$\bullet$] \textbf{Consequences}:
          \begin{itemize}
              \item[$\circ$] Practitioners become distrustful of the explanations given by these tools and avoid using them in high-stakes scenarios. 
          \end{itemize}
          \item[$\bullet$] \textbf{Mitigation strategies}: 
          \begin{itemize}
              \item[$\circ$] Instead of explaining the predictions of a model, make the system more transparent to the user. Custom strategies can be developed for that purpose (e.g. presenting historical performance on similar cases to the user).
          \end{itemize}
        \end{itemize}     
        \\  
        \hline
\end{tabular}
\caption{Overview of explainability issues}\label{soverview:explainability}
\end{table}

\begin{table}
\centering
\begin{tabular}{|c|p{10cm}|}
\hline
Issue Id &  \multicolumn{1}{c|}{Overview of the issues}
\\ \hline
    \multirow{1}{*}[-6.5em]{I3.1}  & \begin{itemize}
          \item[$\bullet$] \textbf{Description}:  Reproducing bugs in an MLSS is difficult because of unstable data sources. A data source is unstable if it returns different values for the same queried record. 
          \item[$\bullet$] \textbf{Cause}: As mentioned in Section \ref{RQ5}, data source providers may forget to serve data with the stability that is required for ML applications. 
          \item[$\bullet$] \textbf{Consequences}:
          \begin{itemize}
              \item[$\circ$] Since bugs can not be reproduced, they tend to be ignored, reducing the trust of stakeholders in the system.
          \end{itemize}
          \item[$\bullet$] \textbf{Mitigation strategies}: 
          \begin{itemize}
              \item[$\circ$] Temporarily save a copy of a record in a database whenever it has been used by a model (for training or inference).
          \end{itemize}
        \end{itemize}     
\\ \hline
    \multirow{1}{*}[-5em]{I3.2}  & \begin{itemize}
          \item[$\bullet$] \textbf{Description}: Debugging data streaming systems (e.g., Hadoop) is difficult because it is difficult to picture what data should look like at each step of the data pipeline. 
          \item[$\bullet$] \textbf{Cause}: The intermediary states of data in a pipeline are not always intelligible for humans (e.g. embeddings). 
        \end{itemize}     
        \\  
        \hline
    \multirow{1}{*}[-5em]{I3.3}  & \begin{itemize}
          \item[$\bullet$] \textbf{Description}: Debugging an MLSS is time-consuming when its data sources are managed by other teams, because it may require inspecting these data sources. 
          \item[$\bullet$] \textbf{Cause}: The priorities of teams managing data sources may not be aligned with the ones of their consumers (i.e. fixing a bug might not be their priority).
          \item[$\bullet$] \textbf{Consequences}:
          \begin{itemize}
              \item[$\circ$] Bugs are not fixed in timely delays.
          \end{itemize}

        \end{itemize}     
        \\  
        \hline
\end{tabular}
\caption{Overview of debuggability issues}\label{soverview:debug}
\end{table}

\begin{table}
\centering
\begin{tabular}{|c|p{10cm}|}
\hline
Issue Id &  \multicolumn{1}{c|}{Overview of the issues}
\\ \hline
    \multirow{1}{*}[-6em]{I4.1}  & \begin{itemize}
          \item[$\bullet$] \textbf{Description}: Training models consume too many resources (e.g., time, computing power, etc.).  
          \item[$\bullet$] \textbf{Cause}: The learning process of large models involves updating numerous parameters over many iterations.
          \item[$\bullet$] \textbf{Consequences}:
          \begin{itemize}
              \item[$\circ$] A lot of money is spent on training the ML models.
              \item[$\circ$] Product development is slowed down.
          \end{itemize}
          \item[$\bullet$] \textbf{Mitigation strategies}: 
          \begin{itemize}
              \item[$\circ$] Parallelize model training across different computers using libraries such as Spark ML. 
          \end{itemize}
        \end{itemize}     
\\ \hline
    \multirow{1}{*}[-8em]{I4.2}  & \begin{itemize}
          \item[$\bullet$] \textbf{Description}: The queries sent to an MLSS are not answered timely (i.e., latency/delay issues). 
          \item[$\bullet$] \textbf{Cause}: Generating a prediction generally involves processing data through a data pipeline and a model. Depending on the implementation, both may consume a significant amount of computational resources.
          \item[$\bullet$] \textbf{Mitigation strategies}: 
          \begin{itemize}
              \item[$\circ$] Use models simpler and smaller models.
              \item[$\circ$] Distribute computing across several machines using frameworks such as PySpark.
              \item[$\circ$] Optimize the data processing pipeline. For example, loops could be replaced by vectorized operations. 
          \end{itemize}
        \end{itemize}     
        \\  
        \hline
    \multirow{1}{*}[-5em]{I4.3}  & \begin{itemize}
          \item[$\bullet$] \textbf{Description}: At inference time, models consume too much memory. 
          \item[$\bullet$] \textbf{Cause}: Larger models may load a lot of parameters into memory to generate predictions. 
          \item[$\bullet$] \textbf{Consequences}:
          \begin{itemize}
              \item[$\circ$] Environments that are memory-constrained can not leverage larger models (which generally have superior prediction performance). 
          \end{itemize}
          \item[$\bullet$] \textbf{Mitigation strategies}: 
          \begin{itemize} 
              \item[$\circ$] Use models designed for memory efficiency.
          \end{itemize}
        \end{itemize}     
        \\  
        \hline
\end{tabular}
\caption{Overview of efficiency issues}\label{soverview:explainability}
\end{table}

\begin{table}
\centering
\begin{tabular}{|c|p{10cm}|}
\hline
Issue Id &  \multicolumn{1}{c|}{Overview of the issues}
\\ \hline
    \multirow{1}{*}[-7em]{I5.1}  & \begin{itemize}
          \item[$\bullet$] \textbf{Description}: Maintaining an MLSS is difficult because there is not enough information describing the data the MLSS consumes. 
          % \item[$\bullet$] \textbf{Cause}: Developers may omit to write documentation for short-term benefits (with potentially long-term costs) \citep{NIPS2015_86df7dcf}.
          \item[$\bullet$] \textbf{Consequences}:
          \begin{itemize}
              \item[$\circ$] Modifying an MLSS becomes slower and error-prone because the knowledge of the data domain gets lost.
          \end{itemize}
          \item[$\bullet$] \textbf{Mitigation strategies}: 
          \begin{itemize}
              \item[$\circ$] Follow rigorous project management methodology where data documentation and understanding play a central role.
              \item[$\circ$] Use a data storage procedure that forces its users to document data when entered into the system.
          \end{itemize}
        \end{itemize}     
\\ \hline
    \multirow{1}{*}[-8em]{I5.2}  & \begin{itemize}
          \item[$\bullet$] \textbf{Description}: Maintaining a model is difficult because there is not enough information describing how the model was generated.  
          % \item[$\bullet$] \textbf{Cause}: Developers may omit to write documentation for short-term benefits (with potentially long-term costs) \citep{NIPS2015_86df7dcf}.
          \item[$\bullet$] \textbf{Consequences}:
          \begin{itemize}
              \item[$\circ$] Without proper documentation, understanding why an MLSS has performance issues can be difficult, leading stakeholders to distrust the MLSS.
              \item[$\circ$] Models may need to be re-developed from scratch if the practitioners who developed them are not available anymore and have not left sufficient documentation.
          \end{itemize}
          \item[$\bullet$] \textbf{Mitigation strategies}: 
          \begin{itemize}
              \item[$\circ$] Use experiment tracking tools, namely MLflow, to automatically document the hyperparameters used to train ML models.
          \end{itemize}
        \end{itemize}     
        \\  
        \hline
    \multirow{1}{*}[-7em]{I5.3}  & \begin{itemize}
          \item[$\bullet$] \textbf{Description}: Managing the dependencies (i.e., software libraries) of an MLSS is challenging and error-prone.  
          \item[$\bullet$] \textbf{Cause}: 
          \begin{itemize}
              \item[$\circ$] The development environment is different from the one in production.
              \item[$\circ$] Code dependencies' versions are not specified, leading to the installation of incompatible libraries.
          \end{itemize}
          \item[$\bullet$] \textbf{Consequences}:
          \begin{itemize}
              \item[$\circ$] Broken environments are deployed in production, hindering the reliability of MLSSs.
          \end{itemize}
          \item[$\bullet$] \textbf{Mitigation strategies}: 
          \begin{itemize}
              \item[$\circ$] Share the same development environment between developers and use that environment in production.
          \end{itemize}
        \end{itemize}     
        \\  
        \hline
\end{tabular}
\caption{Overview of maintainability issues}\label{soverview:explainability}
\end{table}

\begin{table}
\centering
\begin{tabular}{|c|p{10cm}|}
\hline
Issue Id &  \multicolumn{1}{c|}{Overview of the issues}
\\ \hline
    \multirow{1}{*}[-6em]{I6.1}  & \begin{itemize}
          \item[$\bullet$] \textbf{Description}: Having a reliable model is difficult because of concept or data drift.  
          \item[$\bullet$] \textbf{Cause}: The cause of data drift depends on the domain of the MLSS.
          \item[$\bullet$] \textbf{Consequences}:
          \begin{itemize}
              \item[$\circ$] MLSSs have poor prediction performance.
              \item[$\circ$] MLSSs are unable to deliver predictions.
          \end{itemize}
          \item[$\bullet$] \textbf{Mitigation strategies}: 
          \begin{itemize}
              \item[$\circ$] Automatically retrain ML models on the latest data.
              \item[$\circ$] Define a systematic data collection process to avoid annotation drift.
          \end{itemize}
        \end{itemize}     
\\ \hline
    \multirow{1}{*}[-10em]{I6.2}  & \begin{itemize}
          \item[$\bullet$] \textbf{Description}: Having a reliable model is difficult because of external data providers.  
          \item[$\bullet$] \textbf{Cause}: The data providers change how they serve data without taking into consideration how it impacts downstream MLSS. 
          \item[$\bullet$] \textbf{Consequences}:
          \begin{itemize}
              \item[$\circ$] Same as issue I2.1
          \end{itemize}
          \item[$\bullet$] \textbf{Mitigation strategies}: 
          \begin{itemize}
              \item[$\circ$] Before using a data source, assess its reliability and, if it is not reliable, avoid using it.
              \item[$\circ$] Fetch the same information from different data providers to have data redundancy.
              \item[$\circ$] Monitor the data sources to address any issue in the MLSS as fast as possible.
              \item[$\circ$] In case of an issue with a data source used at inference, inform the user that the quality of the model’s predictions will be degraded.
              \item[$\circ$] In case of an issue with data used for training, roll back to a previous version of the model.
          \end{itemize}
        \end{itemize}     
        \\  
        \hline
    \multirow{1}{*}[-5em]{I6.3}  & \begin{itemize}
          \item[$\bullet$] \textbf{Description}: Having a reliable MLSS is difficult because of the data pipelines which are brittle and have technical debt.  
          % \item[$\bullet$] \textbf{Cause}: The cause of brittleness depends on many factors. Technical debt is introduced in code when designs with short-term benefits but long-term costs are left in the code \citep{NIPS2015_86df7dcf}.
          \item[$\bullet$] \textbf{Consequences}:
          \begin{itemize}
              \item[$\circ$] MLSSs have poor prediction performance.
              \item[$\circ$] MLSSs are unable to deliver predictions because of software crashes.  
          \end{itemize}
        \end{itemize}     
        \\  
        \hline
\end{tabular}
\caption{Overview of reliability issues}\label{soverview:explainability}
\end{table}

\end{document}

%% file: title_page.tex
\title{Quality Issues in Machine Learning Software Systems %\thanks{Grants or other notes
%about the article that should go on the front page should be
%placed here. General acknowledgments should be placed at the end of the article.}
}
% \subtitle{Do you have a subtitle?\\ If so, write it here}

%\titlerunning{Short form of title}        % if too long for running head

\author{Pierre-Olivier Côté \and
        Amin Nikanjam \and
        Rached Bouchoucha \and
        Ilan Basta \and
        Mouna Abidi \and
        Foutse Khomh
        %etc.
}

\authorrunning{Côté et al.} % if too long for running head

%\authorrunning{Short form of author list} % if too long for running head

\institute{Pierre-Olivier Côté \and Amin Nikanjam \and Rached Bouchoucha \and Ilan Basta \and Mouna Abidi \and Foutse Khomh \at 
    Polytechnique Montréal, Québec, Canada \\
    \email{\{pierre-olivier.cote, amin.nikanjam, rached.bouchoucha, ilan.basta, mouna.abidi, foutse.khomh\}@polymtl.ca} \\
    Corresponding author's e-mail address: pierre-olivier.cote@polymtl.ca
}

\date{Received: date TODO / Accepted: date}
% The correct dates will be entered by the editor

\maketitle

\begin{abstract}
\textbf{Context}: An increasing demand is observed in various domains to employ Machine Learning (ML) for solving complex problems. ML models are implemented as software components and deployed in Machine Learning Software Systems (MLSSs). \textbf{Problem}: There is a strong need for ensuring the serving quality of MLSSs. False or poor decisions of such systems can lead to malfunction of other systems, significant financial losses, or even threats to human life. The quality assurance of MLSSs is considered a challenging task and currently is a hot research topic. \textbf{Objective}: This paper aims to investigate the characteristics of real quality issues in MLSSs from the viewpoint of practitioners. This empirical study aims to identify a catalog of quality issues in MLSSs. \textbf{Method}: We conduct a set of interviews with practitioners/experts, to gather insights about their experience and practices when dealing with quality issues. We validate the identified quality issues via a survey with ML practitioners. \textbf{Results}: Based on the content of 37 interviews, we identified 18 recurring quality issues and 21 strategies to mitigate them. For each identified issue, we describe the causes and consequences according to the practitioners' experience. \textbf{Conclusion}: We believe the catalog of issues developed in this study will allow the community to develop efficient quality assurance tools for ML models and MLSSs. A replication package of our study is available on our public GitHub repository\footnote{\url{https://github.com/poclecoqq/quality_issues_in_MLSSs}}.

\keywords{Machine Learning based Software Systems \and Quality Assurance \and Quality issues\and Interview \and MLOps \and Machine Learning}
\subclass{MSC 68T05}
\end{abstract}